\documentclass[preprint,12pt]{elsarticle}
\pdfoutput=1
\usepackage{graphicx}
\usepackage{subfig}
\usepackage{threeparttable}
\usepackage{amsmath,amsfonts,amsthm, amssymb} 
\usepackage{lineno,hyperref}
\modulolinenumbers[5]
\usepackage{multirow}
\usepackage{booktabs}
\usepackage{caption} 
\usepackage[dvipsnames]{xcolor}
\newcommand{\vect}[1]{\boldsymbol{#1}}

\newcommand{\tb}{\textbf}
\makeatletter
\def\mathcolor#1#{\@mathcolor{#1}}
\def\@mathcolor#1#2#3{%
  \protect\leavevmode
  \begingroup
    \color#1{#2}#3%
  \endgroup
}
\makeatother

\makeatletter
\def\ps@pprintTitle{%
 \let\@oddhead\@empty
 \let\@evenhead\@empty
 \def\@oddfoot{}%
 \let\@evenfoot\@oddfoot}
\makeatother

\begin{document}

\begin{frontmatter}


\title{A Robust Approach for Stability Analysis of Complex Flows Using Navier-Stokes Solvers}
\author{Rajesh Ranjan\corref{mycorrespondingauthor}}
\ead{ranjan.25@osu.edu}
\author{S. Unnikrishnan, Datta Gaitonde\corref{}}
\address{Department of Mechanical \& Aerospace Engineering\\ The Ohio State University, Columbus, OH, 43210}
\cortext[mycorrespondingauthor]{Corresponding author}

\begin{abstract}
 Global stability modes of flows provide significant insight into their dynamics. 
Direct methods to obtain these modes are restricted by the daunting sizes and complexity of Jacobians encountered in general three-dimensional flows.
Jacobian-free approaches have greatly alleviated the required computational burden. 
Particularly, the most common Arnoldi-based methods obtain the desired subset of the eigenmodes by considering Jacobian-vector products to create a smaller iterative subspace, instead of working with the Jacobian itself.
However, operations such as orthonormalization and shift-and-invert transformation of matrices with appropriate shift guesses  can introduce computational and parameter-dependent costs that inhibit their routine application to general three-dimensional flowfields. 
Further, in time-stepper type approaches, where the linearized perturbation snapshots directly obtained from an aerodynamic code are treated as Jacobian-vector products, the inversion operation  necessitates use of approximate iterative linear solvers with several parameters.
The present work addresses these limitations by proposing and implementing a robust, generalizable  approach to extract the principal global modes, suited for curvilinear coordinates as well as the effects of compressibility.
Accurate linearized perturbation snapshots are obtained using high-order schemes by leveraging the same non-linear Navier-Stokes code as used to obtain the basic state by appropriately constraining the equations using a body-force.
The forcing includes not only that required to obtain the products, but also to ensure that the basic state does not drift.
It is shown that with random impulse forcing, dynamic mode decomposition (DMD) of the subspace formed by these products yields the desired physically meaningful modes, when appropriately scaled.
The leading eigenmodes are thus obtained without spurious modes or the need for an iterative procedure.
Further since  orthonormalization is not required, large subspaces can be processed to capture converged low frequency or stationary modes.
The validity and versatility of the method is demonstrated with numerous examples encompassing essential elements expected in realistic flows, such as compressibility  effects and complicated domains requiring general curvilinear meshes. 
Favorable qualitative as well as quantitative comparisons with Arnoldi-based method, complemented with substantial savings in computational resources show the potential of current approach for  relatively complex flows. 
\end{abstract}

\end{frontmatter}

\section{Introduction}
\label{sec:intro}
Linear stability analysis (LSA) explores the evolution of small disturbances in an equilibrium state to yield the stability dynamics of a flow. 
The stability modes are obtained either through an initial value problem (IVP), or, in the asymptotic limit, from an eigenvalue problem (EVP). 
The former is non-modal and can yield algebraic transient growth, while the latter is  modal and provides exponential growth with harmonic time-dependent perturbations.
The present study is concerned with the latter objective, which formulates the EVP from the linearized Navier-Stokes (NS) equations, which may be written in terms of evolution of peturbations $\vect{Q'}$ in compact matrix form as:
\begin{eqnarray}\label{eq:lnse}
\displaystyle \frac{\partial \vect{Q'}}{\partial t} = \vect{A}\vect{Q'}
\end{eqnarray}

At the fundamental level, the linear stability dynamics of a flow can be characterized using eigenvalues of the Jacobian matrix $\displaystyle \vect{A} = \frac{\partial \vect{F}}{\partial \vect{Q}}\Bigr|_{\substack{\vect{Q}=\vect{\bar{Q}}}}$, where $\vect{\bar{Q}}$ is the basic state and $\vect{F}$ is the flux vector in the corresponding non-linear equations. 
The size of the square matrix $\vect{A}$, whose each dimension scales with the total number of degrees of freedom, is a major limiting factor, which restricts routine application primarily to low-dimensional, often canonical problems. 

At the beginning, LSA was used successfully to generate key insights into locally one-dimensional homogeneous systems such as parallel flows. However, advances in computational power as well as numerical algorithms have increasingly extended their scope to global two- and three-dimensional flows. 
Interested reader can refer to the review articles by Theofilis\citep{theofilis2003advances,theofilis2011global}, and the references therein for a detailed list of applications. 

The number of applications, where stability study can offer valuable insights, is growing enormously. 
However, stability analyses of complex configurations, whose basic state determinations are facilitated by curvilinear configurations, as well as those involving compressibility effects, are also relatively uncommon. 
Iterative approaches have shown considerable success in overcoming some of the difficulties associated with such studies such as prohibitively large matrix size, since they avoid directly solving the full EVP.
Most common iterative approaches are based on subspace iteration methods such as Arnoldi \cite{arnoldi1951principle} or Krylov-Schur \cite{stewart2002krylov}, in which approximations to eigenvalues and eigenvectors are obtained by constructing an orthonormal basis of the Krylov subspace. 
The methods converge to eigenvalues which are of largest modulus, and are especially effective when they are very distinct from each other.
Since these approaches solve only for a part of eigenspectrum, they may effectively target modes that are either dynamically less relevant or spurious. 

Stability studies typically seek modes which first become unstable.
Representing the eigenvalue of a mode as $\omega_r + i \omega_i$, modes whose imaginary components ($\omega_i$) representing growth rates first become positive by crossing the real zero axis are particularly desired. 
This is typically achieved by using a spectral transformation of the Jacobian $\vect{A}$ using shift-and-invert (or, two-parameter Cayley)  such that  eigenvalues of the new matrix $\displaystyle \vect{A^{ST}} = \frac{\vect{I}}{(\vect{A}-\sigma_s \vect{I})}$ (or, $\displaystyle \vect{A^{ST}} =   \frac{(\vect{A}-\mu \vect{I})}{(\vect{A}-\sigma_s \vect{I})}$) converge close to a parameter $\sigma_s$. 
Here $\vect{I}$ is an identity matrix and $\sigma_s$ is a complex quantity whose value should be close to the expected frequency ($\omega_r$) and growth rate ($\omega_i$) of the desired mode. These approaches are thus very efficient provided $\sigma_s$ is correctly assigned. 
This can present a major limitation in using such techniques to study the stability dynamics of flows which have complex physics and a good estimate of $\sigma_s$ is \textit{a priori} not available. 

Efficient implementation of the shift-and-invert method is greatly enabled by the explicit availability of the Jacobian matrix. 
This approach, designated `matrix-forming'\cite{theofilis2000globally}, has been successfully used to determine the stability dynamics of many flows such as lid-driven cavity \cite{bergamo2015compressible}.
The approach becomes less practical for realistic three-dimensional flows, where either the Jacobian may not be readily available or is intractable, such as in flows with higher-order schemes and intricate boundary conditions, or can become very large.

A significant advantage in the determination of eigenmodes accrues from considering the linearized perturbation snapshots educed from a code solving LNSE (Eq. \ref{eq:lnse}) as Jacobian-vector products $\vect{A}\vect{Q'}$. 
Such efforts, that use the Krylov subspace formed by the evolution of Jacobian-vector products, often called time-stepper \cite{Bagheri2009} or time-stepping\citep{theofilis2011global}, have been successful in many recent applications.
However, these methods incur costs associated with the larger parameter space.
The orthonormalization procedure may place practical constraints on the size of the Krylov space to a few time periods.  
This can result in convergence issues for flows with very-low-frequency instability. 
An example of this limitation was observed in the open cavity stability studies of \citet{Picella2018}, where occurrences of stationary modes were subject to increased uncertainty.
Other convergence issues along with large parameter space may arise because of the application of iterative minimum residual solver required to convert Jacobian-vector product $\vect{A}\vect{Q'}$ obtained directly from the code into a product of spectrally transformed Jacobian and vector ($\vect{A^{ST}}\vect{Q'}$) for creating Krylov subspace. 
Practical considerations for the robustness of iterative methods include choice of discretization scheme and the initial vector from which the products are initiated, possible need for preconditioning of ill-conditioned matrices

For complex flows, the method to get desired linearized perturbation , which form the Krylov subspace, deserves further attention.
One way to obtain these is to directly use an entirely new code solving linearized Navier-Stokes equations. 
However, developing such a solver for practical problems is an arduous task, since each effect or complication in the Navier-Stokes equations must be linearized.
The difficulties arise from the derivation of LNSE in the presence of curvilinear co-ordinate systems, terms associated with compressibility that couple the momentum and energy equations and possible source terms.
From an execution perspective, these may introduce stiffness in the equations, affecting the time-step-size, and require further care in implementing accurate boundary conditions on perturbations.
As such, the effort to develop a separate LNSE solver may be comparable to that of developing the Navier-Stokes solver itself. 

In order to circumvent this issue, some efforts  make use of existing non-linear solvers to obtain the desired linearized perturbation snapshots \cite{mack2010preconditioned, gomez2014three}. These methods usually make use of Jacobian-free Newton-Krylov methods (JFNK, \cite{knoll2004jacobian}), where quasi-linearization of non-linear terms terms is achieved using Fre\`{c}het derivatives. For example,
\begin{equation}\label{eq:frechet}
\displaystyle \vect{A}\vect{Q'} = \frac{\partial \vect{F}}{\partial \vect{Q}}\Bigr|_{\substack{\vect{Q}=\vect{\bar{Q}}}} \vect{Q'} \approx \frac{ \vect{F}({\vect{\bar{Q}}}+\varepsilon \vect{Q'}) - \vect{F}({\vect{\bar{Q}}})} {\varepsilon}
\end{equation}
Where $\varepsilon$ is a small perturbation magnitude, which should be chosen with care to obtain a good approximation of the above derivative without nonlinear contamination, while also avoiding the round-off errors and machine precision limits.
The method has provided significant insights into compressible swept leading edge flow\citep{mack2010preconditioned} and cavity dynamics\citep{gomez2014three} .

In their three-dimensional stability calculations for the  lid-driven cavity using above approach, \citet{gomez2014three} indicate that the numerical error can exceed the perturbation magnitude, which can preclude successful application of the method.
This may be due to the first-order approximation to Jacobian as used in Fre\`{c}het derivatives implementation above, despite the fluxes being usually evaluated to higher-order accuracy in the non-linear simulation component. 
Higher-order approximations of the Taylor series require more flux computations and are generally not employed. 
Other considerations concerning discontinuities and reaction fronts are discussed in \citet{knoll2004jacobian}. \citet{mack2010preconditioned} in their detailed description of the above approach have provided a list of 13 parameters that control the generation of physically meaningful results and the robustness of convergence. Because of large input parameter space along with practical issues associated with orthonormalization and shift-and-invert Arnoldi implementation, the power of the JFNK technique has not yet found routine application in the stability literature.

The current effort seeks pragmatic approach to enable stability studies of complex flows by addressing some of the above concerns. 
Two of the principal challenging steps in existing iterative approaches are: (1) computation of Jacobian-vector products using a non-linear solver with similar order of accuracy as base flow, and (2) extraction of physically meaningful eigenvalues from these products using a cost-effective approach, and without depending on any guess parameter such as shift value.
Both are addressed in this work, in separate steps, by adapting and combining procedures to effectively yield the desired principal stability modes.
The Jacobian-vector products are generated through implicit linearization of the Navier Stokes equations with the same code as the one used to obtain the basic state.
This leverages the features of the non-linear solver, including its advanced high-order spatio-temporal schemes.
An easily formed constraining body force addresses several features that can otherwise affect linearized perturbation evolution using the non-linear Navier-Stokes equations.
In particular, it addresses the fact that the residual of the basic state simulation may not reach machine zero in complex problems with curvilinear coordinates, compressibility and simplified boundary conditions.
It also allows analysis of the properties of turbulent mean flows, such as those obtained from time-averaging of Large-Eddy Simulations (LES) or Reynolds-Averaged Navier-Stokes equations as attempted in some studies including \citet{crouch2007predicting}. 
The linearized response then represents the evolution of the initial disturbance vector, $\vect{Q}_1' = \varepsilon_0 \vect{\bar{Q}}$, where $\varepsilon_0$ is the initial perturbation magnitude. 
The relevant eigenvalues from the subspace are then obtained by using Dynamic Mode Decomposition (DMD) after appropriate scaling. 

Since the two steps \textit{i.e.,} generation of the subspace  and  extraction of eigenmodes are decoupled, convergence can be verified by varying the number and sampling frequencies of Jacobian-vector operations as well as the dependence, if any, on the initial vector in the subspace. 
Likewise, the current approach avoids expensive orthonormalization and shift-and-invert transformations, allowing for the examination of very low frequency phenomena by using large subspace.
Also the absence of normalization between time-steps reduces the sensitivity of the results to the perturbation magnitude $\varepsilon$, as the imposed perturbations no longer need to be scaled to unity at every time-step. 
This eliminates the requirement of adapting $\varepsilon$ every time step, usually followed in time-stepper approach \citep{gomez2014three}. 

The effectiveness of the current approach is demonstrated in a range of applications.
These span dimensionality (and thus eigenvalue matrix size), compressibility, and complexity of domains (curvilinear coordinates). 
For future reference, the different applications considered are shown in table \ref{tab:applications}.            
\begin{table*}
\captionsetup{justification=centering}
\caption{Problems Considered. 2DLDC=Two-dimensional Lid-driven Cavity. 3DLDC=Three-dimensional Lid-driven Cavity. CYL2D=Flow past a Cylinder. NACA2D=NACA0015 Airfoil at Stalled Conditions.}
\begin{center}
\label{tab:applications}
\def\arraystretch{1.6}
\begin{tabular}{|p{2.1cm}|p{1.8cm}|p{1.8cm}|p{8.0cm}|}
\toprule
\multirow{2}{*}{\tb{Case}} &  \multirow{2}{*}{\tb{Re}} &\multirow{2}{*}{\tb{M}}  &\multirow{2}{*}{\tb{Remarks}}    \\
& &   & \\
\midrule
\tb{2DLDC}  & 11200 & 0.95 & Compressible two-dimensional lid-driven cavity. Uniform Cartesian mesh.  Qualitative and quantitative comparison with the Arnoldi approach. \\
\hline
\tb{3DLDC} & 2100 & 0.1  & Fully three-dimensional cavity with end walls. Post-bifurcation eigenvalue prediction in high degree-of-freedom systems \\
\hline
\tb{CYL2D} & 40, 50 & 0.1 & Pre- and post-bifurcation analysis of first wake instability in a circular  cylinder. Curvilinear application to cylindrical stretched meshes\\
\hline
\tb{NACA2D} & 200 & 0.1, 0.5  & NACA0015 airfoil at high angle of incidence. General curvilinear mesh system. Effects of compressibility. \\
\bottomrule
\end{tabular}
\end{center}
\end{table*}
The cases chosen are motivated to highlight the pragmatic advantages of the method in handling realistic geometries and flows with relative ease. A detailed comparison with Arnoldi approach is presented for a compressible two-dimensional lid-driven case, along with the description of requirement of computational resources in each approach. 
For other flows considered, comparisons are made with the stability results obtained using existing methods, wherever possible. 


\section{Mode Extraction Technique}\label{sec:mfp}
The proposed approach falls in the general category of time-stepper methods.
The two main components, the method to generate Jacobian-vector products, and the subsequent post-processing step that extracts the principal stability modes from the subspace, are now described

\subsection{Generation of Jacobian-Vector products}
The desired Jacobian-vector products ($\vect{A}\vect{Q'}$), which essentially are linearized peturbation snapshots, are obtained by subjecting a modified non-linear Navier-Stokes solver to an initial perturbation field. 
The method, designated as the Navier-Stokes-based mean flow perturbation (NS-MFP) technique, is based on the approach that is described in references\cite{touber2009large,bhaumik2018verification}) but is employed here in a different manner and purpose.
The current derivation, more suited for creating a subspace to extract the stability modes  is summarized below.

Here, the basic state,  $\vect{\bar{Q}}$, is obtained with the non-linear Navier-Stokes code as usual.
In the general case \textit{i.e.,} with curvilinear coordinates and compressible effects, the residual may not reach machine zero.
Likewise, if an asymptotic unsteady state is reached, then the basic state may be considered to be the time-averaged solution, whose residual to the Navier-Stokes operator is also non-zero.
In such cases, the modes are representative of the linearized dynamics of turbulent mean flows, and define the larger turbulent scales in the flow \citep{crighton1976stability,tam1995supersonic}.

To account for these properties of basic states, a single application of the NS operator is used to obtain the change, $B_f$ to the basic state:
\begin{equation}\label{eqn:R}
B_f = \frac{\partial  \vect{\bar{Q}}}{\partial t} - \vect{F}( \vect{\bar{Q}})
\end{equation}
The discretizations used to obtain the derivatives on the right side of Eqn.~\ref{eqn:R}, leverage the underlying non-linear Navier-Stokes solver, and are described in Section~\ref{sec:method}.
If the basic state is an exact solution of the discretized non-linear Navier-Stokes equations, then $B_f$ is zero.
However, for reasons above, in practical cases, $B_f$ is a finite spatially-varying quantity.

The perturbation forcing is then imposed on $\vect{\bar{Q}}$ and the solution is marched forward as usual.
However, at the end of each step, the stored valued $B_f$ is subtracted from the change in the solution vector.
This ensures that any changes in the basic state due to the NS operator are explicitly removed.
The governing equations of NS-MFP are thus the non-linear equations with body force $B_f$:
\begin{equation}
\frac{\partial \vect{Q}}{\partial t} = \vect{F}(\vect{Q}) + B_f
\end{equation}
which may be written as:
\begin{equation}
\frac{\partial ( \vect{\bar{Q}}+\vect{Q'})}{\partial t} = \vect{F}( \vect{\bar{Q}}+\vect{Q'}) + B_f
\end{equation}
Applying Taylor series expansion around the base flow $ \vect{\bar{Q}}$ and rearranging yields:
\begin{equation}
\displaystyle \frac{\partial \vect{Q'}}{\partial t} + \left(\frac{\partial \vect{\bar{Q}}}{\partial t} - \vect{F}(\vect{\bar{Q}}) - B_f \right) = \vect{Q'} \frac{\partial \vect{F}}{\partial \vect{Q}} + \vect{Q'}^2 \frac{\partial^2 \vect{F}}{\partial \vect{Q}^2} + H.O.T.
\end{equation}
Using relation \ref{eqn:R}, the second term on left hand side becomes zero.  
Thus,
 \begin{equation}
\displaystyle \frac{\partial \vect{Q'}}{\partial t} =  \vect{Q'} \frac{\partial \vect{F}}{\partial \vect{Q}} + \vect{Q'}^2 \frac{\partial^2 \vect{F}}{\partial \vect{Q}^2} + H.O.T.
\end{equation}
As in LNSE, if the imposed perturbation $\vect{Q'}$ is chosen to be sufficiently small compared to $\vect{\bar{Q}}$,  flow as required to ensure linearity, second- and higher- order terms may be neglected, and the above equation reduces to the known LNSE equation given as:
\begin{equation}
\displaystyle \frac{\partial \vect{Q'}}{\partial t} =  \vect{Q'} \frac{\partial \vect{F}(\vect{\bar{Q}})}{\partial \vect{Q}} \end{equation}

\begin{figure} 
\centering
\includegraphics[width=.85\textwidth]{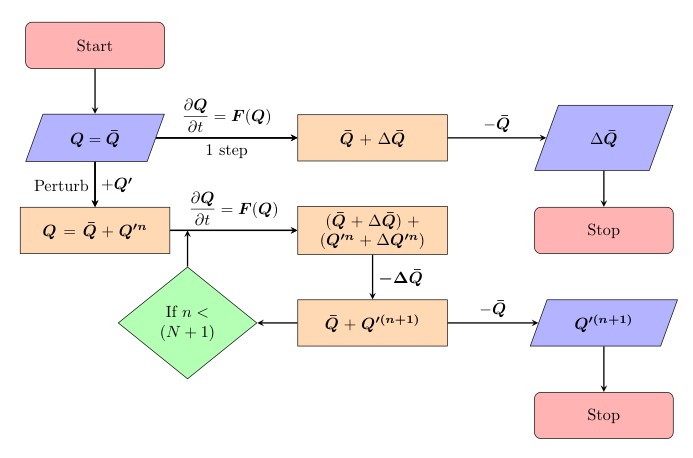}
\caption{Navier-Stokes-based mean flow perturbation (NS-MFP) approach for generating Jacobian-vector products. Here $\vect{Q'}^{(n+1)} = \vect{A}\vect{Q'}^n$. $\Delta \vect{{\bar{Q}}}$ is the change in basic state and is comparable to body-force $B_f$ defined in the description.}
\label{fig:flowcharts}
\end{figure}
The practical implementation of the above approach in a given non-linear code is shown through flowchart in 
Figure~\ref{fig:flowcharts}.
Implicit linearization of NS equations is thus achieved by using the change in the basic state due to the Navier-Stokes operator as a restraining body force and, as further discussed below, employing a small $\vect{Q'}$.  Present approach gives flexibility to apply boundary conditions on either the basic state or perturbations, which are discussed in Section~\ref{sec:bc}.  
\subsubsection{Forcing characteristics}
The acquisition of stability modes is crucially governed by the forcing or the nature of perturbation superimposed on the base flow.
Depending on the objectives, the forcing is imposed  locally or globally in an impulsive or continuous fashion.
A general form for the imposed perturbation can be written as:
 \begin{equation}
\displaystyle \vect{Q'}(\vect{x},t) = \varepsilon ~\phi(\vect{x}) \psi(t)
\end{equation}
where $\varepsilon$ is the amplitude (or perturbation magnitude), and $\phi$ and $\psi$ are the spatial and temporal distributions of the perturbations.
If a harmonic perturbation is considered, it may be represented as:
\begin{equation}
\displaystyle \vect{Q'}(\vect{x},t) = \varepsilon ~\phi(\vect{x}) \Re(e^{i~2\pi \omega~t})
\end{equation}
where $\omega$ is the excitation frequency and $\phi(\vect{x})$ is the corresponding mode shape. 
$\Re(.)$ represents real part of the complex quantity enclosed within the parenthesis. 
\citet{bhaumik2018verification}, for example, used the inviscid stability solution of the entropy layer evolving over a flat plate with a blunt leading edge with different forcing amplitudes, $\varepsilon$.
NS-MFP then yielded an excellent match for the growth rates of each tested mode.
For jet noise studies, they used localized but continuous excitation at different frequencies on a mean turbulent basic state to obtain the larger unsteady scales in the jet, and hence the jet noise characteristics.

Our goal however is to obtain the linear stability modes, rather than responses at specific frequencies, for arbitrary flow fields.
For this, since the leading frequency of the flow is unknown \text{a priori} a suitable forcing is comprised of random broadband noise which has no preferred frequency. 
This noise may be applied to the entire volume or a localized region in  the computational domain based on the nature of stability to be explored. 
These perturbations are then evolved without additional forcing \textit{i.e.,} the forcing is applied as an impulse at $t = 0$. 
Thus, the form of the perturbation forcing is:
\begin{equation}
\displaystyle \vect{Q'}(\vect{x},t) = \varepsilon_0 ~ \vect{Q} ~\mathrm{RAND}(\vect{x}) \delta(t)
\end{equation}
where $\mathrm{RAND}(\vect{x})$ generates a random normalized distribution in space in the range $[-1~~ 1]$, which is scaled with the free-stream value of the base flow variable $\vect{Q}$ and initial perturbation magnitude $\varepsilon_0$. $\delta(t)$ is the temporal distribution of the disturbance; in this case the Dirac-delta function specifies impulse forcing. 
The choice of initial random noise as forcing also aids in establishing the linear independence of snapshots necessary for stability mode extraction, as discussed below.

Different values of $\varepsilon_0$ have been examined for each of the test problems below, to discern values needed to ensure linearity, which is established through tests for proportionality between input and output.
Although the value of $\varepsilon_0$ should be evaluated for each  problem,  for all cases considered here, $\varepsilon_0 = 10^{-5}$ was found to be appropriate to maintain linearity while avoiding machine precision limits. Because we use compact higher-order schemes as described later, the accuracy of the scheme does not become a limiting factor in assigning the value of $\varepsilon_0$.

The final choice concerns the variables to which the impulse forcing should be applied.
Flows containing shear ease this choice, since an initial impulse in any one variable rapidly excites responses in the other flow variables, \textit{i.e.,} the background basic state rapidly manifests the primary structures in all variables, as governed by the dynamical equations.
Thus, any variable suffices; for the cases considered below, only pressure perturbation is imposed on the base flow. 

\subsection{Extraction of Eigenvectors from Subspace}\label{sec:dmd}
Eigenvalues and eigenmodes are usually extracted from snapshot series using iterative methods such as Arnoldi iterations.
As noted earlier, in the present method we avoid orthonormalization or shift-and-invert transformations.
Specifically, we employ Dynamic Mode Decomposition (DMD; \cite{schmid2010dynamic}), which serves as a variant of a standard Arnoldi method, but does not require the orthonormalization between time-steps. 
DMD performs data-driven dimensionality reduction and requires no additional inputs, once the input subspace consisting of snapshots is chosen. 
The decomposition of this subspace gives modes, in which each of them are characterized by a single frequency. 
Thus, $N$ snapshots separated by a constant sampling time $\Delta t$, may be assembled in subspace $\displaystyle \mathbb{Q}(\vect{x},t)$, which can be represented as summation of DMD modes:
\begin{eqnarray} \label{eq:dmd}
  \displaystyle \mathbb{Q}(\vect{x},t) = \sum_{n=1}^N  \Phi_n (\vect{x}) e^{\omega_n t}
\end{eqnarray} 
where $\Phi(\vect{x})$ and $\omega (= \omega_r + i\omega_i)$ represent spatial support and the complex frequency of the mode.
These modes approximate the modes of a Koopman operator using a finite set of data, the latter of which is a linear, infinite-dimensional operator representing non-linear, finite dimensional dynamics \citep{bagheri2013koopman, chen2012variants}. 
Further, if the data are obtained from a linear dynamical system, DMD then yields eigenvalues and eigenvectors of that system \cite{schmid2010dynamic}. 

Since its introduction to fluid dynamics, DMD has been used extensively on experimental and numerical data to extract dynamically dominant modes.
Its capability to extract stability modes though mentioned in the original work\citep{schmid2010dynamic}, this approach is not widely used in the stability literature. The primary reason for this is that while low-dimensional canonical flows (with a good estimate of shift value) can be solved using Arnoldi approach, there has not been many studies of high degree-of-freedom systems where DMD may have an advantage. Further, in the absence of an accurate high-order method to generate the Jacobian-vector products and thus the input DMD subspace, the results from this approach may be a crude approximation to what would have been obtained from classical approach. The high order method proposed in the earlier section thus make the input data reliable for DMD, and therefore brings the stability results much closer to those obtained using classical approach as shown through several examples in the following sections.  

The key features that determine the suitability of DMD for stability mode extraction purposes are further outlined here.
The frequency-based extraction of modes makes DMD suitable for stability analysis, if the quality of the source data conform to the requirements of DMD \citep{duke2012error}.  
Two main requirements for minimum residual error in DMD are: (1) the snapshots should be a result of a linear mapping and (2) initial linearly independent snapshots should be followed by linear dependence. 
Both of these are naturally accounted for in NS-MFP solutions if  sufficiently many snapshots are used at a suitable sampling frequency. 
A large number of snapshots ensures the required eventual linear dependence among consecutive realizations: this constraint in DMD is also a requirement for modal asymptotic analysis.

In the current approach, DMD is used as a post-processing tool without direct coupling to the procedure used to obtain the Jacobian-vector products.   
Once the decomposition is done, dynamic modes are then ranked based on their contribution to the first vector in the subspace \citep{chen2012variants}. 
This ranking brings the desired unstable modes which dominate the dynamics of flow during bifurcation among the leading modes. 
Further, secondary and higher-order stability modes, if present in the flow, which affect the dynamics are also extracted among the dominant ones.

Of the many variants of DMD, we follow the singular value decomposition (SVD)-based approach as discussed in \citet{schmid2011applications}.
In this, a low-dimensional matrix is constructed through a similarity transformation of the subspace formed using perturbation snapshots.   
The eigenvalue $\omega$ of the Jacobian $\vect{A}$ is related to the eigenvalue $\lambda$ of the DMD matrix through $\omega = \mathrm{log}{\lambda}/\Delta t$. It should noted that here $\omega_r$ and $\omega_i$ represent here growth rate and associated frequency, exactly opposite to the convention usually followed in the classical stability analysis. Therefore, for future reference we designate the linear frequency and growth rate of a mode as $\omega$ and $\sigma$ respectively. The circular frequency $\omega_c$ is related to linear frequency as $\omega_c =2\pi \omega$. It should be mentioned here that since the flows are simulated using scales non-dimensionalized with characteristic length and velocity, the linear frequency obtained represents the Strouhal number ($St$).  

The resolved frequencies in DMD depend on the Nyquist criterion and the total sampling time, while numerical convergence can be ensured by successively increasing the number of snapshots.
The perturbation amplitude may likewise be varied: however, if the disturbance decay/growth is linear as in the present case, the results are insensitive to this choice.

Another key choice necessary to the use of DMD for stability studies is the flow variable to be subjected to the DMD procedure.
Typically, like Arnoldi-based methods, snapshots with all the primitive flow variables can be used to create the subspace.
However, since DMD is decoupled from the process of generation of snapshots, one can use only those primitive or even derived variables in the subspace which influence the dynamics most.
For most of the two-dimensional analyses performed included in this study, the vorticity variable alone was found to be sufficient to extract the leading stability dynamics. 
The stability studies, performed with all five primitive variables (as in classical approach) as well as with only with velocity variables yield similar results. 
Therefore, the current approach may save upto 80\% of the computational memory compared to classical approach during the mode extraction process if appropriate variable is chosen. This saving may prove crucial for high dimensional systems. 
For the three-dimensional lid-driven cavity analysis included in this study, all three velocity variables were considered for  extraction of modes so as to include any spanwise effects. 

\section{Numerical Schemes}\label{sec:method}
As noted earlier, a strength of the method to generate Jacobian-vector products is that it uses the same framework as that for obtaining the basic state, thus leveraging the treatment of mesh metrics and spatio-temporal discretization schemes.
The key step include a pre-calculation the body force term to arrest any changes in the basic state.
The details are now summarized for completeness.

\subsection{Governing Equations}
The compressible Navier-Stokes equations are solved in non-dimensional form on a curvilinear $(\xi,\eta,\zeta)$-coordinate system:
\begin{equation} 
\frac{\partial}{\partial \tau} \biggl(\frac{\vect{Q}}{J}\biggl) = -\biggl[\biggl(\frac{\partial \vect{F_i}}
{\partial \xi} + \frac{\partial \vect{G_i}}{\partial \eta} + \frac{\partial \vect{H_i}}
{\partial \zeta}\biggl) + 
\frac{1}{Re} \biggl(\frac{\partial \vect{F_v}}{\partial \xi} + \frac{\partial \vect{G_v}}{\partial \eta} 
+ \frac{\partial \vect{H_v}}{\partial \zeta}\biggl) \biggl]                
\label{NSE}
\end{equation}      
\noindent where, $\vect{Q}=[\rho,\rho u,\rho v, \rho w, \rho E]^T$ denotes the solution vector defined in terms of the fluid density $\rho$, Cartesian velocity components $(u,v,w)$ and total specific internal energy $E={T}/{(\gamma-1)M^2}+(u^2+v^2+w^2)/2$. 
Here, $M$ is the reference Mach number, $\gamma$ is the ratio of the specific heats and $T$ is the fluid temperature. 
The ideal gas law connects fluid-pressure $p$ to $\rho$ and $T$ as $p=\rho T/{\gamma M^2}$. 
Sutherland's law is used to express fluid viscosity $\mu$ as a function of temperature $T$.  
$J=\partial{(\xi,\eta,\zeta,\tau)}/\partial{(x,y,z,t)}$ is the Jacobian of the transformation from the Cartesian $(x,y,z)$  to the curvilinear $(\xi,\eta,\zeta)$-coordinate system. 
The inviscid and viscous fluxes in ($\xi, \eta, \zeta$)-directions are represented in Eq.~(\ref{NSE}) by $(\vect{F_i}, \;\vect{G_i}, \;\vect{H_i})$ and $(\vect{F_v}, \;\vect{G_v}, \;\vect{H_v})$, respectively. 

\subsection{Numerical Discretization}
A sixth-order compact difference scheme is employed to compute the flux terms of the governing equations. 
An eighth-order implicit filter with $\alpha = 0.45$ is used for numerical stability. 
A detailed discussion of the numerical algorithm used for the simulations can be found in \cite{gaitonde1998high}.  
The time-stepping is performed using nonlinearly stable third order Runge-Kutta method proposed by ~\citet{shu1988efficient}.  

\subsection{Boundary Conditions}\label{sec:bc}
The cases considered include wall and farfield boundaries. 
For the basic state, no-slip conditions are applied at the wall, while wall-normal gradients are set to zero for pressure and density. 
For far field conditions, as necessary for the cylinder case, characteristics-based boundary conditions are used.

Significant flexibility on boundary condition implementation is available when generating the Jacobian-vector products, since they are derived by advancing the total variable (basic state plus perturbation).
Thus, the boundary conditions may be applied on either the total flow or only on perturbations. 
For the studies presented below, wall boundary conditions are applied to the perturbations. 
At farfield boundaries, as in the flow past the cylinder as well as the airfoil, characteristic boundary conditions are enforced on the total flow.

\section{Results}\label{sec:cases}
We now consider four applications which include the use of simple and curvilinear meshes as well as compressibility to highlight the features of the proposed approach.

\subsection{Compressible Two-Dimensional Lid Driven Cavity (2DLDC)}\label{sec:ldc}
The stability of a lid-driven cavity flow in a box configuration with a moving top wall has been an area of active research for many decades \citep{kuhlmann1997flow, theofilis2004viscous}.
Central to many of these studies is the mechanism of first Poincar\'{e}-Andronov-Hopf bifurcation at which a steady flow turns periodic. 
Since the changes in the dynamics of the flow with Reynolds number are very evident in LDC, this flow is routinely used as a classic benchmark problem in stability analysis. 

Although numerous studies have explored the stability characteristics of incompressible flows, their compressible counterparts have not received as much attention. 
The compressible governing equations are more cumbersome to linearize and the subsequent discretization of the governing equations also requires more care. 
A key conclusion from recent compressible studies \cite{bergamo2015compressible, canuto2015two, Ohmichi2017compressibility} concerns the stabilizing effects of compressibility.

In order to assess the accuracy of our approach, we present stability results for a two-dimensional cavity with regularized moving top wall as suggested in \citet{shen1991hopf}. 
The problem is relatively simple, with well-defined boundary conditions and amenable to discretization with uniform Cartesian  grids.
The stability results are presented for a high Mach number flow ($M=0.95$) at a corresponding critical Reynolds number $Re~=~11,200$ \citep{Ohmichi2017compressibility}. 
. 

\begin{figure}
\centering
\includegraphics[width=0.9\textwidth]{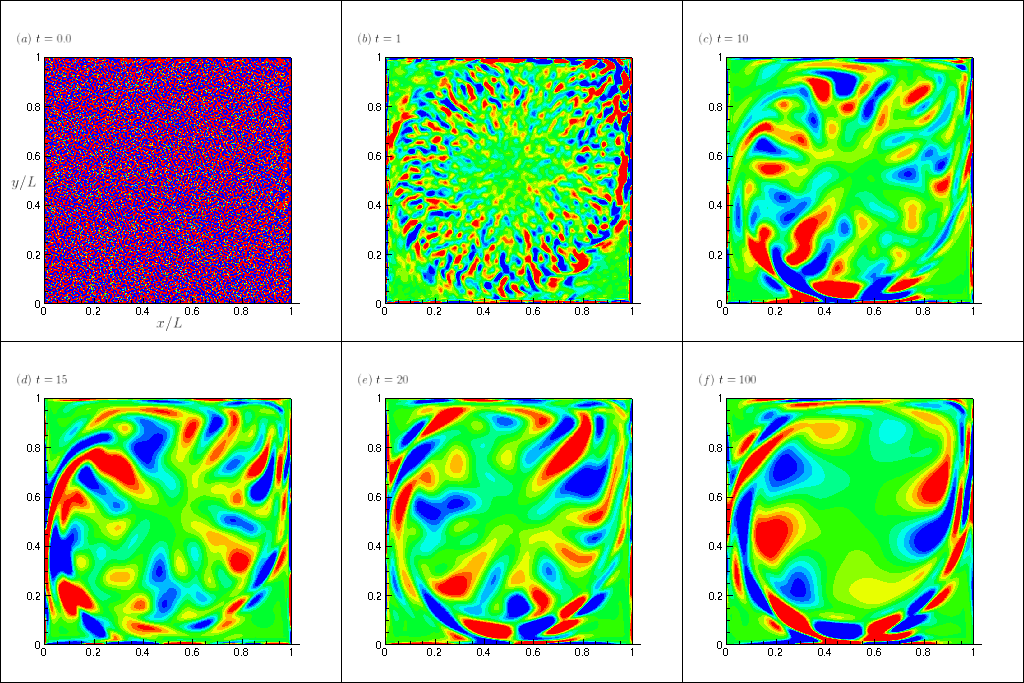}
\caption{Linear evolution of perturbations using NS-MFP (2DLDC). Perturbation vorticity contours are plotted at 11 levels between $-2\times10^{-6}$ and $2\times10^{-6}$.}
\label{fig:Re6000_perturb}
\end{figure}
Both the base flow as well as Jacobian-vector products are generated 
on a uniform grid with size $301\times301$ and a non-dimensional time step of $\Delta t = 5\times10^{-4}$.  
To form the Jacobian-vector products, an initial random impulse perturbation was applied only to the pressure field with an amplitude of $10^{-5}\bar{P}$.
Transient stages of evolution of the perturbation are shown in Fig. \ref{fig:Re6000_perturb} with snapshots of vorticity at various time instants. 
The disturbances evolve slowly under the influence of the background base flow to form coherent spatial structures. 

The snapshot data from the NS-MFP procedure are sampled at every $50$ iterations \textit{i.e.,} at an effective $\Delta t$ of $0.025$.  
A total of $10,000$ snapshots are collected for DMD post-processing.
With 36 processors, the procedure requires less than one hour of wall-clock time.
This relatively large number of snapshots is far more than necessary, but is chosen to establish convergence and linearity tests.
Specifically, the subspace size was varied from 500 snapshots to 5000 snapshots, with different starting snapshots and sampling frequencies.

The results obtained from current approach are compared with those from a BiGlobal (two inhomogeneous directions) matrix-forming approach using Arnoldi. 
The BiGlobal code is an in-house code written in MATLAB (version R2018b) exploiting sparse linear algebra capabilities.
The code solves eigenvalue problem formulated by retaining all the terms in linearized compressible Navier-Stokes equations.
Several differentiation schemes, including second and higher order central differencing as well as those based on Chebyshev and Hermite polynomials are available for user to choose. The two explicit finite-difference schemes with 3-point and 5-point stencils which denote second- and fourth order central differencing respectively, are designated as FD2 and FD4 respectively. The effect of choosing a Chebyshev differentiation scheme (designated as CHEB) for this flow is also presented. For two-dimensional stability study presented below, the spanwise wavenumber is set to zero to enable comparison. 

\begin{figure}
\centering
\includegraphics[width=1.1\textwidth]{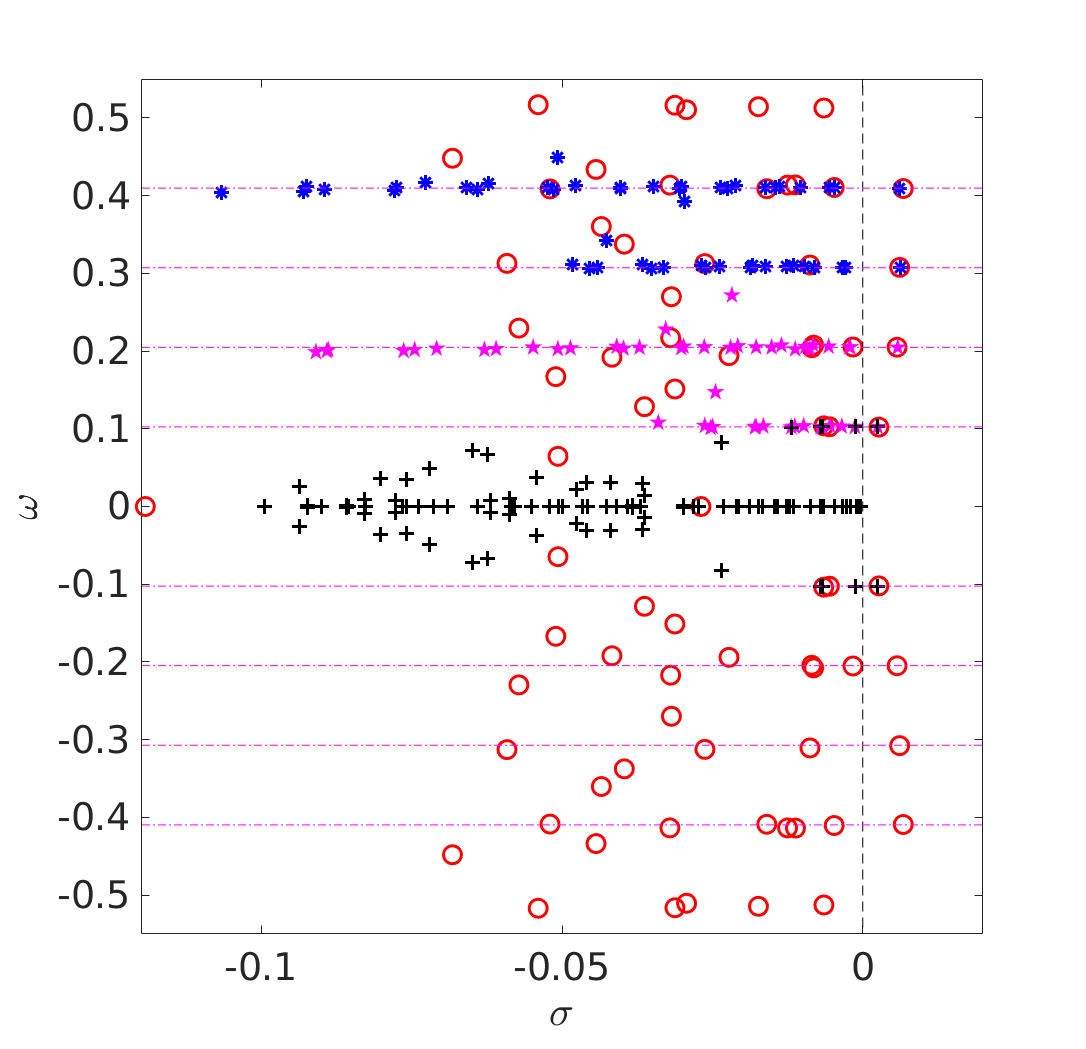}
\caption{Eigenspectrum obtained using current approach compared with those using Arnoldi with different shift $\sigma_s$ values (2DLDC). \textcolor{red}{O}- Present approach with 4000 snapshots. $+$ - Arnoldi with $\sigma_s/(2\pi) = 0, m=100$. \mbox{\large{$\textcolor{magenta}{\star}$}} -  Arnoldi with $\sigma_s/(2\pi) = 0.2, m=50$, \mbox{\large{$\textcolor{blue}{\ast}$}} - Arnoldi with $\sigma_s/(2\pi) = 0.4, m=50$. See tables  \ref{tab:arnoldi2} and \ref{tab:arnoldi} respectively for computational costs and convergence details.}
\label{fig:Biglobal}
\end{figure}
Figure~\ref{fig:Biglobal} displays the eigenvalues obtained using current approach as well as from Arnoldi. 
In the current approach, only vorticity variable was used during mode extraction process with 4000 snapshots in the subspace.
The Krylov subspace size $m$ in Arnoldi was varied between 50 and 100, so was the shift parameter $\sigma_s$ in order to extract the relevant part of eigenspectrum. 
The results shown here are with FD2 scheme, which require least computational resources.
Both the Stability results were obtained on $301\times301$ grid, which was also used to obtain the basic state.

The eigenspectrum obtained using current approach shows that the flow is unstable at this Reynolds number. 
We obtain four unstable modes ($\sigma > 0$) with frequencies, designated for reference as $\omega_1, \omega_2, \omega_3$ and $\omega_4$, which are harmonics of the fundamental frequency $\omega_1 = 0.1023$.
The harmonic nature of the modes is anticipated because of the trivial wall boundary conditions as also observed in the case of low Mach number stability studies \citep{ohmichi2016assessment}. 
Further, the unstable eigenvalues relate to the instability of unstable rotating inviscid Couette flow as reported by \citet{chiba1998global}, whose linear frequency is given as: $\omega_k = k\Omega/2\pi$, where $\Omega \simeq 0.7$ is the angular velocity of the flow and $k$ is the wavenumber in the circumferential direction. 
The context of Couette flow instability in current LDC study will be become clear when the structures of the modes will be discussed later.

The effect of Mach number is seen in the frequencies of the modes obtained. 
When compared to the frequency reported in the literature for incompressible or low Mach number studies \cite{ohmichi2016assessment}, around 7\% drop in $\omega_k$ is observed in the current compressible study.  
This decrease in  frequency seems to be the effect of compressibility, as such dependence with Mach number was earlier observed in the case of circular cylinder\citep{canuto2015two}.
Further discussion on the effects of compressibility will be out of scope for the present work.
 
We return to examining the results obtained with Arnoldi in Fig.~\ref{fig:Biglobal} at the default shift value ($\sigma_s = 0$), we note that only $\omega_1$ mode was obtained with subspace size $m=100$. We did not find any unstable mode with $m=50$. Increasing the subspace size beyond $m=100$ may enable capturing of higher unstable modes, but that study had impractical computational requirements with the current grid size.

When a correct estimate for shift $\sigma_s$ was used for targeting a particular instability mode, we were able to capture that mode with subspace size $m=50$ or even lower. For example, in Fig.~\ref{fig:Biglobal}, we see modes corresponding to frequencies  $\omega_1$, and $\omega_2$ extracted with $\sigma_s/2\pi = 0.2$. With $\sigma_s/2\pi = 0.4$, we notice most unstable modes corresponding to frequencies  $\omega_3$, and $\omega_4$ successfully recovered. Because of positive shift values used, modes with negative frequency in the oscillatory pair are not recovered. In terms of quantitative comparison between Arnoldi results and those obtained with current approach, both the frequencies as well as growth rates are remarkably close to each other. Further, even though we have used 4000 snapshots for computing the modes shown in the Figure, we were able to extract the leading modes with only 1000 snapshots.  

\begin{table*}
     \centering
\begin{threeparttable}
\begin{center}
\caption{Comparison of computational performance for grid size $301\times301$ (2DLDC)}
     \label{tab:arnoldi2}
\def\arraystretch{1.6}
\begin{tabular}{p{1.7cm}|p{2.5cm}|p{2.2cm}|p{1.9cm}}
\toprule
\tb{Method}  & \tb{Subspace Size} ($m$)  & \tb{Memory (GB)} &  \tb{Time (sec)} \\
\midrule
Arnoldi  & 100 & 66.3 & 8073\\
Arnoldi & 50 & 12 & 381 \\
Current Approach\tnote{1} & 4000 & 18 & 450 \\
\bottomrule
\end{tabular}
\begin{tablenotes}
\item[1] cost of generating subspace for mode extraction not included.
\end{tablenotes}
\end{center}
\end{threeparttable}
\end{table*}
\begin{table*}
\centering
\caption{Stability Results for the Most Unstable Mode (2DLDC)}
\begin{center}
\label{tab:arnoldi}
\def\arraystretch{1.6}
\begin{tabular}{p{1.7cm}|p{2.1cm}|p{2.2cm}|p{2.1cm}|p{1.5cm}}
\toprule
\multirow{2}{*}{\tb{Method}} & \tb{Frequency} &\tb{Growth rate}  & \tb{Grid Size (BiGlobal)}  & \tb{Diff. Scheme}    \\
& ($\omega$)& ($\sigma$) &     &    \\
\midrule
\tb{Current} & 0.409100349  & 0.006740765 & $301\times301$  & Compact  \\
\hline
\multirow{7}{*}{\tb{Arnoldi}} & 0.408537052  & 0.005419862  & $101\times101$  & \multirow{3}{*}{FD2}  \\
 & 0.409013721 & 0.005918495 & $201\times201$   & \\
 & 0.409084385 & 0.006054891 & $301\times301$   & \\
\cline{2-5}
  & 0.409364976  & 0.005409836 & $101\times101$ & \multirow{2}{*}{FD4} \\
  & 0.409157597  & 0.005727509 & $201\times201$   & \\
  \cline{2-5}
  & 0.367846862  & 0.035953738 & $51\times51$  & \multirow{2}{*}{CHEB} \\
  & 0.412583725  & 0.008125815 & $101\times101$   & \\
\bottomrule
\end{tabular}
\end{center}
\end{table*}
Now, we compare the computational requirements for both the methods which give practical insights in using these approaches. The computational resources required for both mode extraction processes are given in Table \ref{tab:arnoldi2}.  
Both the studies are performed on a Linux desktop workstation with 64 Intel\textsuperscript{\textregistered} Xeon\textsuperscript{\textregistered} E7-4809 v4 (20M Cache, 2.10 GHz) processors, and 256 GB RAM. It may be noticed that in the Arnoldi approach, even for moderate grid size of $301\times301$ and subspace size of 100, a huge amount of system memory is required due to orthonormalization and matrix inversion operations involved. The compute time taken is also high compared to other two studies listed in the table, however, that is not that much a limiting factor than memory. When the subspace size in Arnoldi is reduced to $m=50$, the requirement for system memory drastically goes down. However, small subspace size is relevant only if a correct estimate of the shift value is available, else the desired part of eigenspectrum may not be recovered as already observed in Fig. ~\ref{fig:Biglobal}. 

The requirement of computational resources in the current approach is very affordable even with 4000 snapshots in the input subspace as can be seen from table \ref{tab:arnoldi2}. The memory and time requirements given in this table do not include the resources required for generating the subspace as this step is decoupled from mode extraction process. However, the resources required for generating the subspace is only a fraction of that required to generate the basic state, and more importantly it can be performed in parallel without any memory limitations. It should be mentioned here that in the case of Arnoldi, employing a time-stepper approach (\cite{gomez2014three}) instead of matrix-forming as used here may result in considerable drop in memory usage. Nonetheless, the ability of current approach to extract all unstable eigenvalues with reasonable computational power and without a guess shift parameter is remarkable. 

For further quantitative analysis, we shift our attention to the least stable mode at $\omega_4 = 0.4091$. Table \ref{tab:arnoldi} quantifies the results with the current and Arnoldi methods, the latter with three difference schemes as well as different grid sizes in stability calculations after interpolating the base flow. 
\begin{figure}
\centering
\subfloat[$\omega_1 \simeq 0.1023, \sigma_1 \simeq 0.0025, k=1$] 
{\includegraphics[width=.3\textwidth]{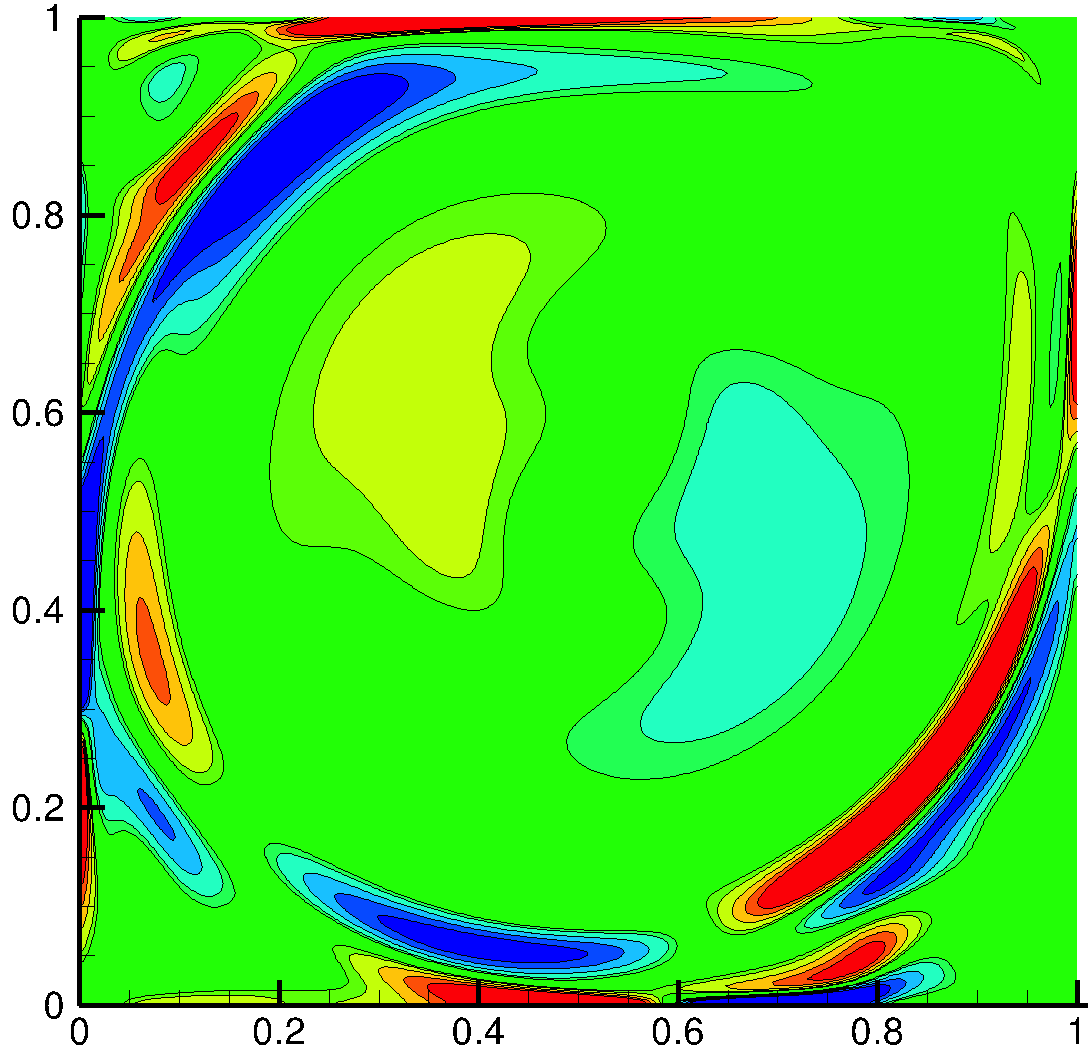}\qquad
\includegraphics[width=.3\textwidth]{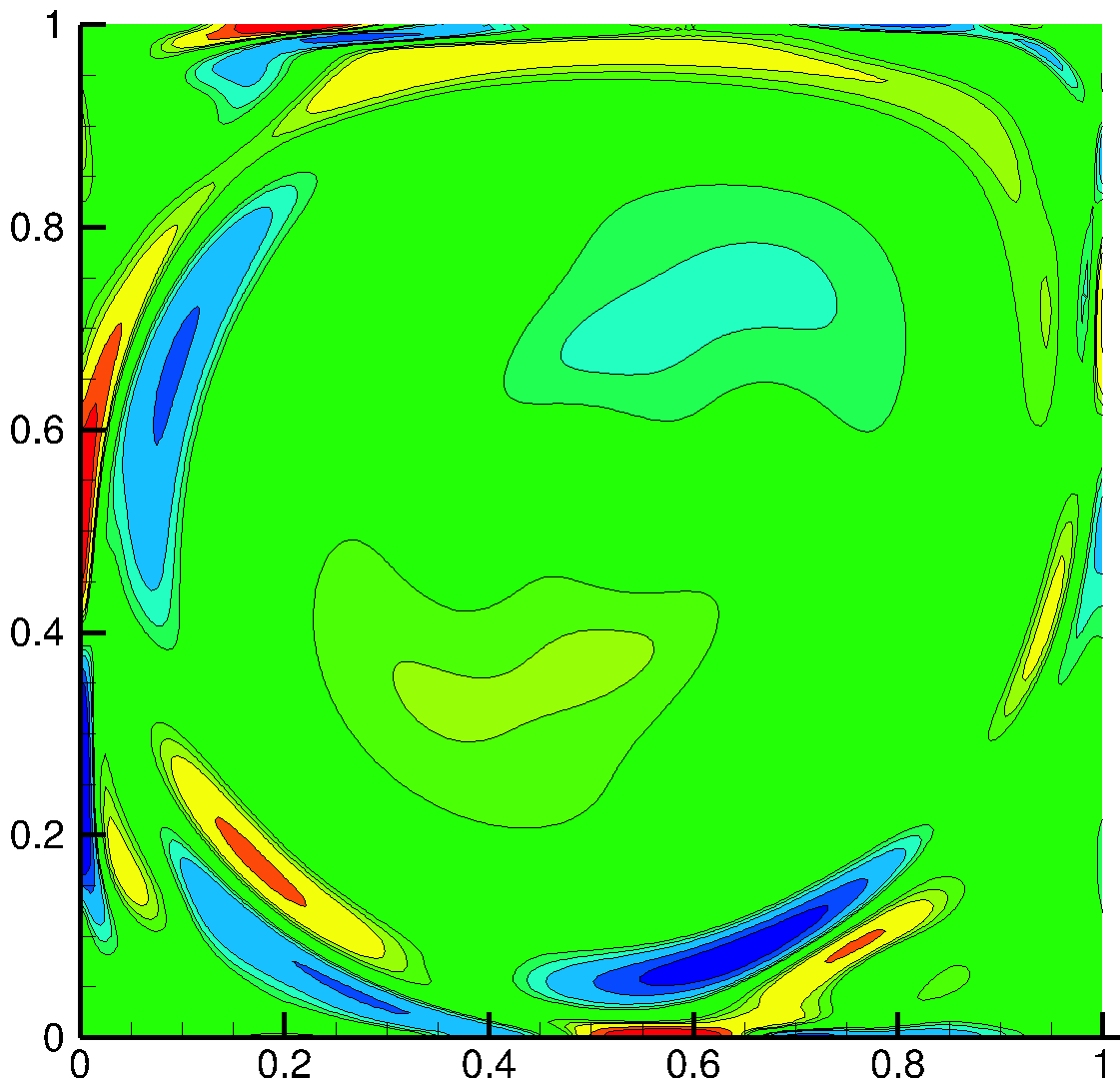}}\\ \vspace{-1em}
\subfloat[$\omega_2 \simeq 0.2046, \sigma_2 \simeq 0.0058, k=2$] 
{\includegraphics[width=.3\textwidth]{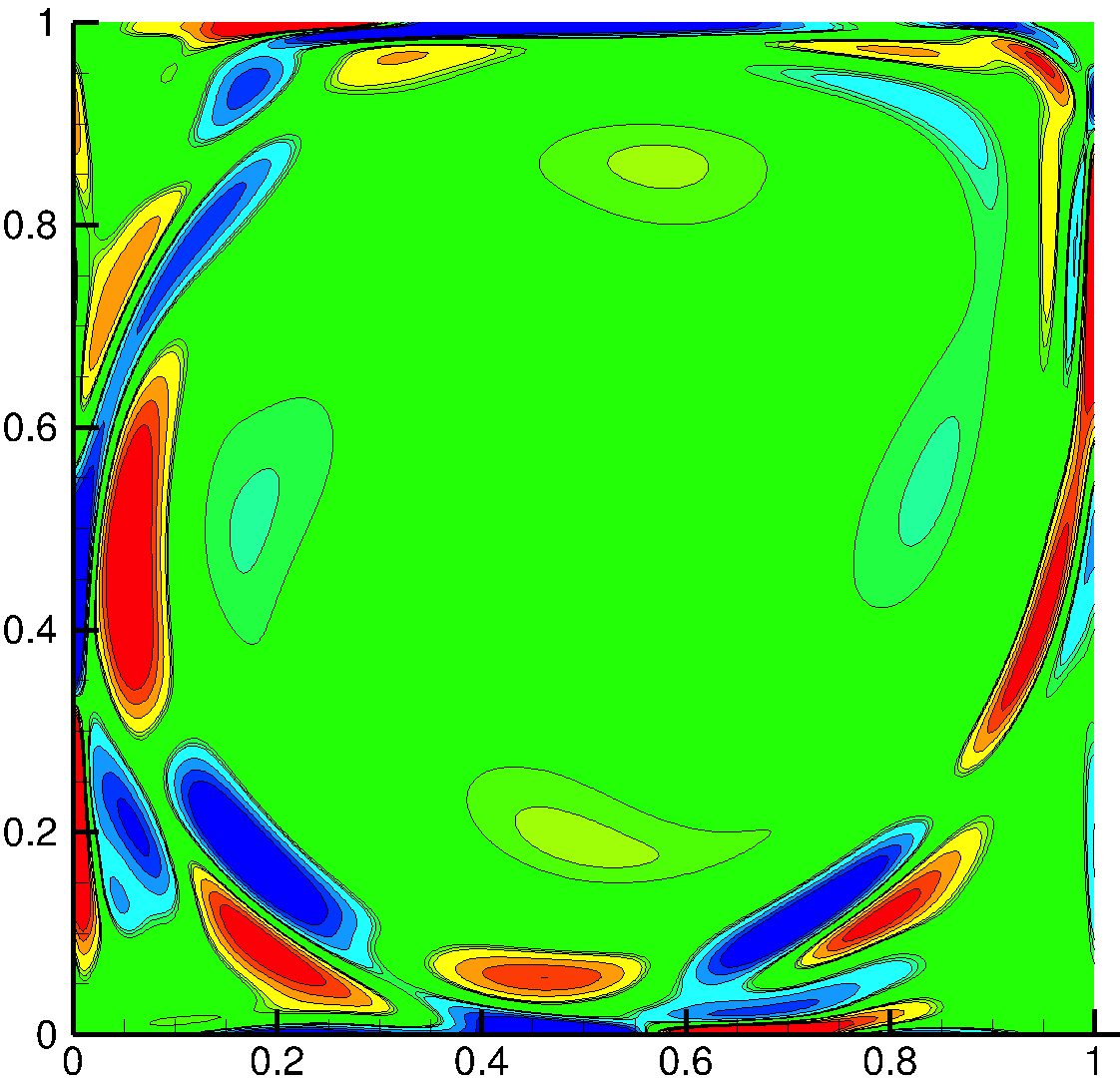}\qquad
\includegraphics[width=.3\textwidth]{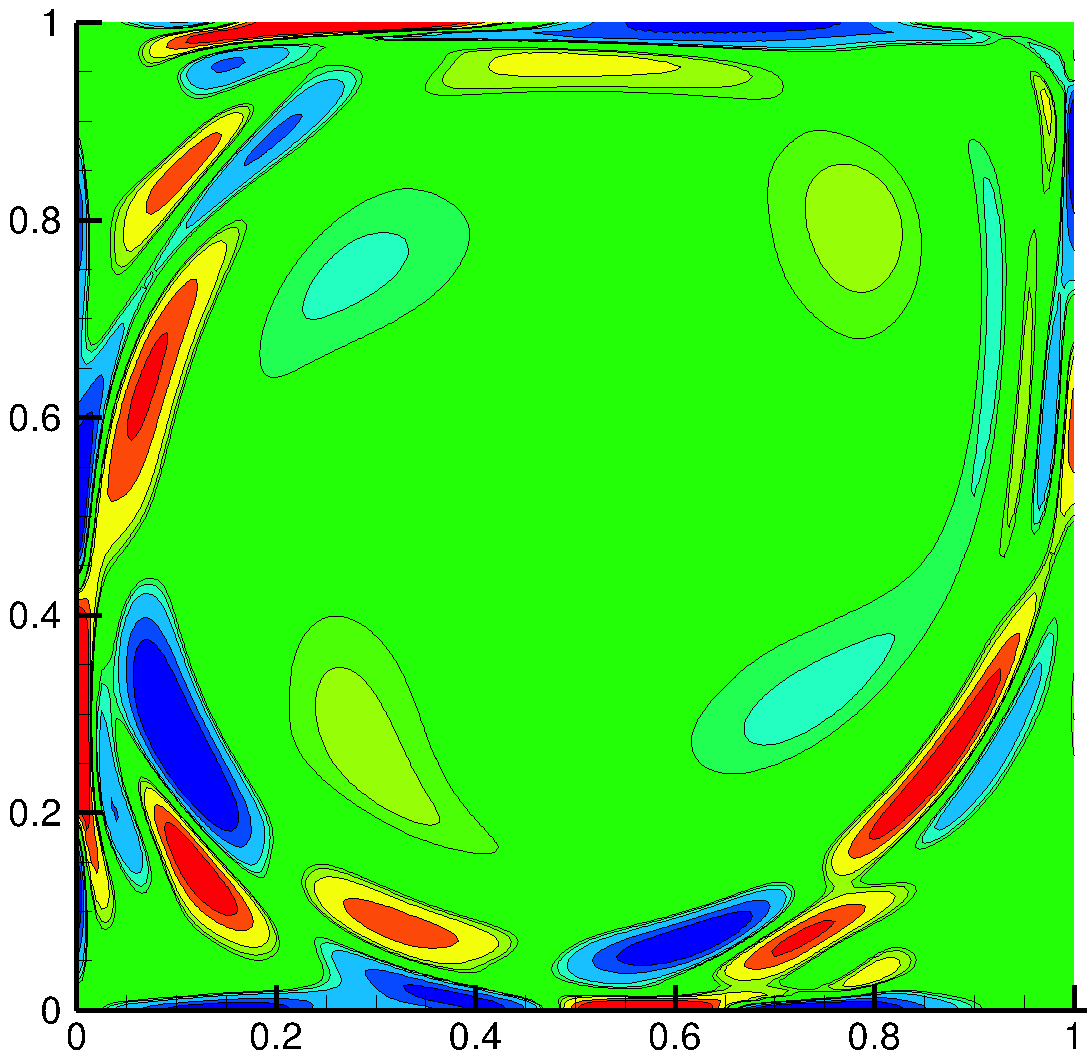}}\\ \vspace{-1em}
\subfloat[$\omega_3 \simeq 0.3069, \sigma_3 \simeq 0.0062, k=3$] 
{\includegraphics[width=.3\textwidth]{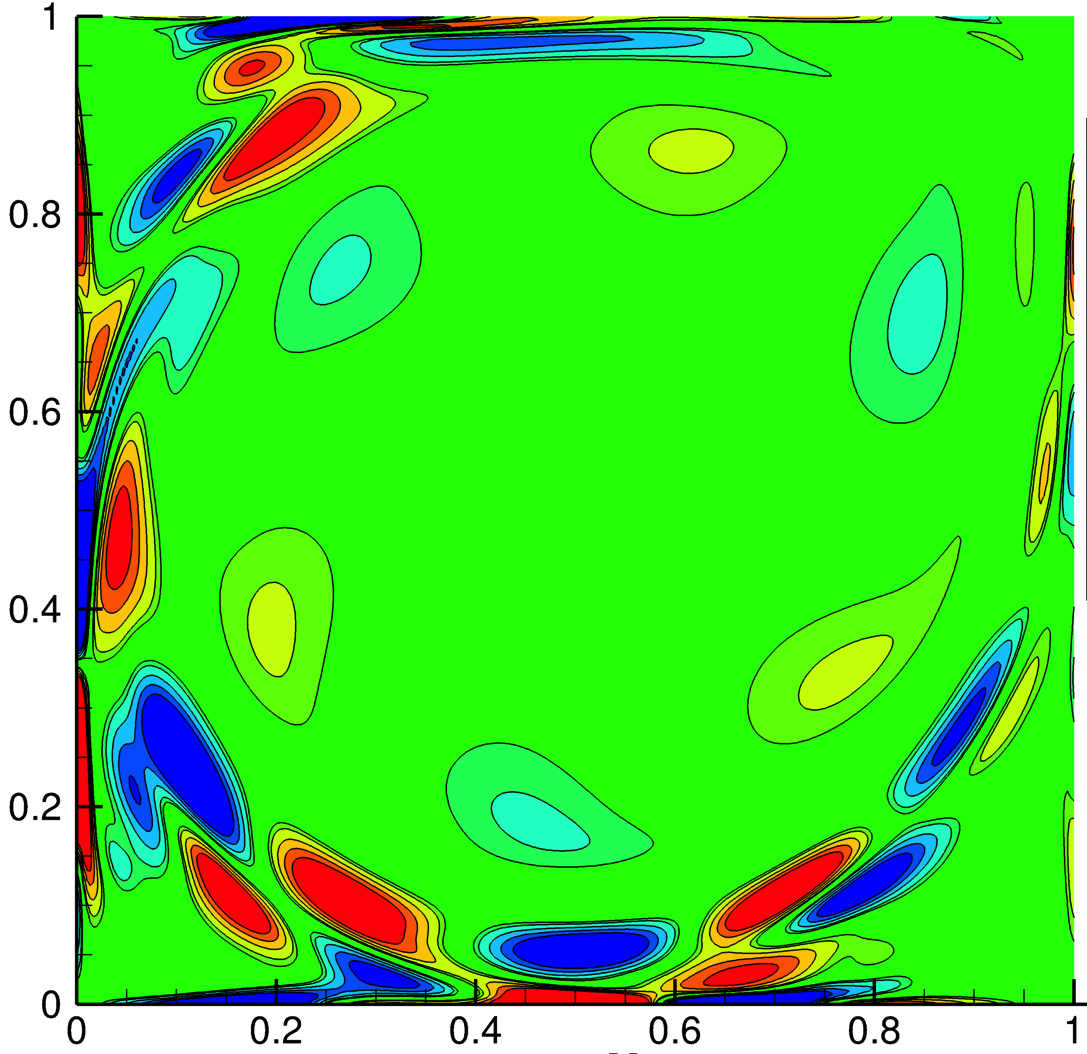}\qquad
\includegraphics[width=.3\textwidth]{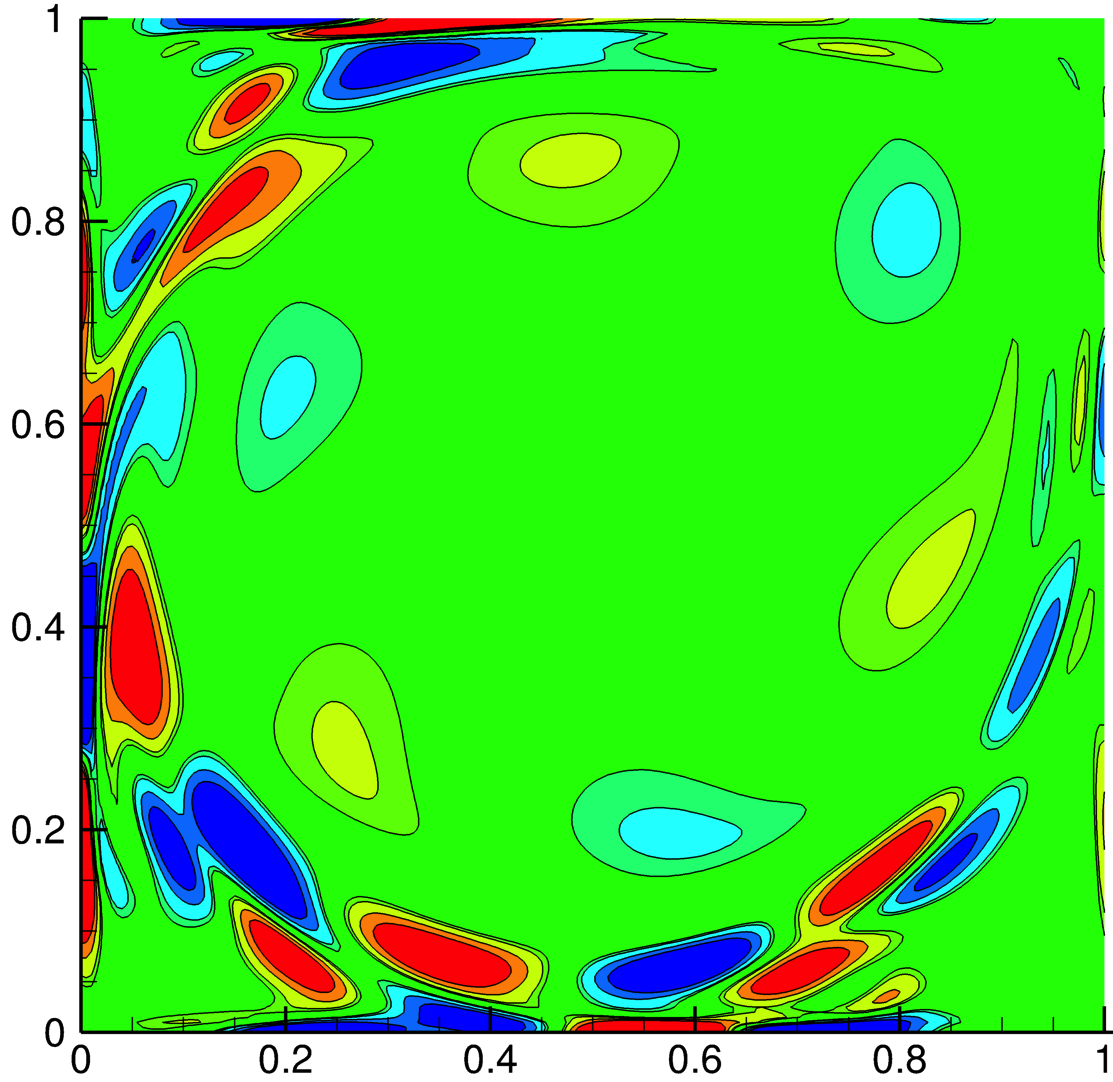}}\\ \vspace{-1em}
\subfloat[$\omega_4 \simeq 0.4091, \sigma_4 \simeq 0.0067, k=4$] 
{\includegraphics[width=.3\textwidth]{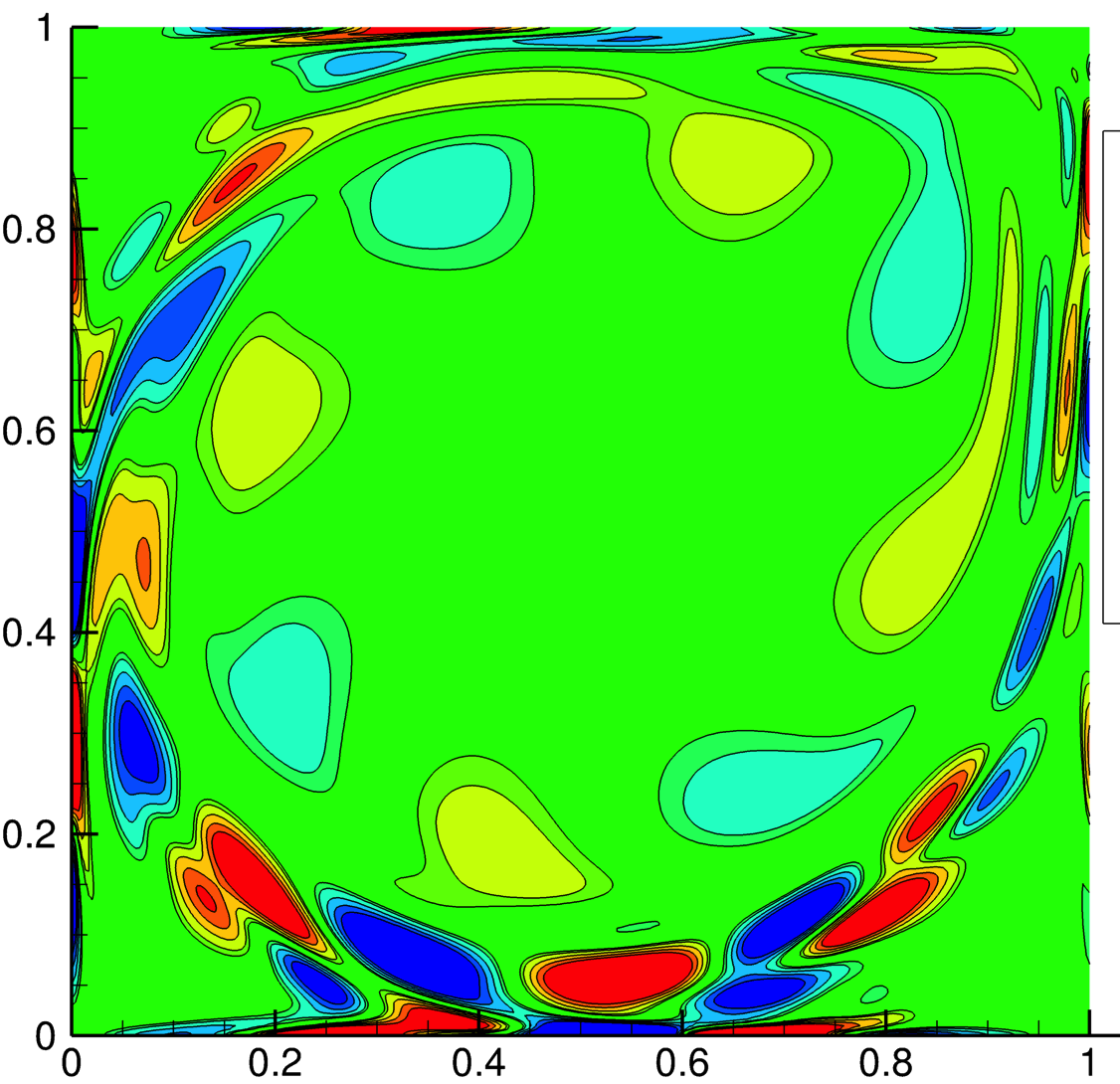}\qquad
\includegraphics[width=.3\textwidth]{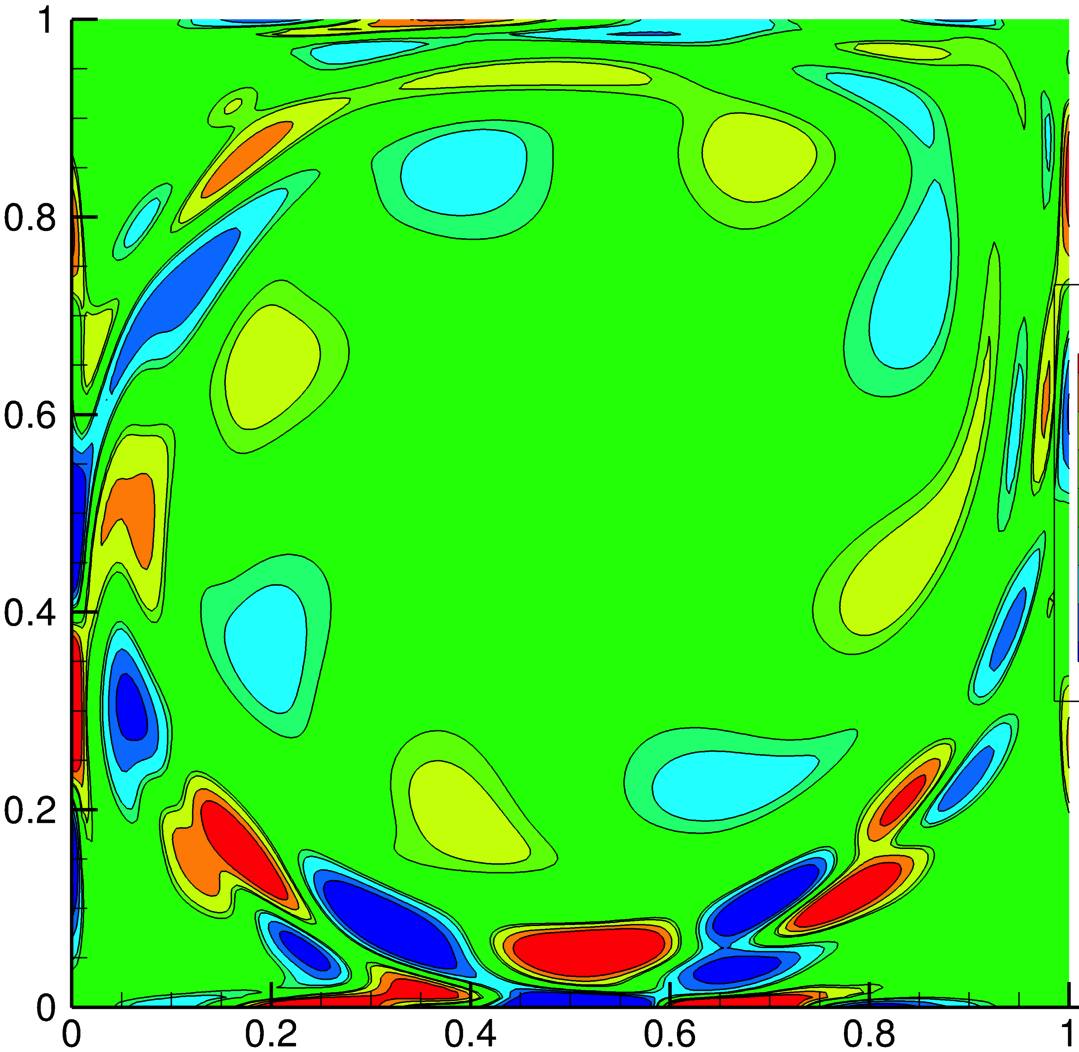}}
\caption{Structures of least Stable modes shown through perturbation vorticity contours. The frequency of a mode is given by $\omega_k = k\Omega/2\pi$, where $k$ is the circumferential wavenumber given by the number of pair of vortices in the primary vortex region. Left: Current Approach. Right: BiGlobal (2DLDC). }
\label{fig:Biglobal2}
\end{figure}
The first thing to note is that with finite difference schemes in Arnoldi, the values of frequencies and growth rates converge with less than 5\% difference between $201\times201$ and $301\times301$ grids for FD2 case. 
On comparing these values with those obtained using current approach, we note that the results are same upto three decimal places. 

An interesting observation from Table \ref{tab:arnoldi} is the performance of the Chebyshev method in extracting the eigenvalues. The approach is much more expensive than explicit finite-difference schemes, because of its low sparsity, and is usually considered more accurate in flows where gradients of interest are concentrated near the wall. 
However, in this case, for both $51\times51$ and $101\times101$ grids, the eigenvalues obtained are further away than those obtained from FD2 and FD4. 
This indicates that the stability dynamics may not dominated by the wall shear region.

In order to examine this aspect, we look at the structures of the four unstable modes in Fig.~\ref{fig:Biglobal2}. 
A quick glance shows that the instability is primarily concentrated in two regions for all the modes: (1) primary vortex region at the core, and (2) circumferential shear-flow region around the vortex. 
Though the shear-flow region is dominating at this $Re$, the frequencies of the modes are prescribed by the instability in the primary vortex.
This primary vortex region can be considered similar to a inviscid rotating Couette flow as studied in \cite{chiba1998global} which has same angular velocity $\Omega$ as found from DNS \citep{ohmichi2016assessment}.
Thus, much like Couette flow instability, the frequency of the instability modes in 2DLDC is given by a relation between angular velocity and circumferential wavenumber, $\omega_k = k\Omega/2\pi$ as mentioned earlier. 
The number of pair of counter-rotating vortices in the primary vortex region in each mode as shown in Fig.~\ref{fig:Biglobal2} give the circumferential wavenumber $k$. 

Finally, we comment on the structures of modes obtained using two methods. 
There is striking resemblance of the mode shapes between the two methods with key features captured in each case. 
The number of lobes in the core region as well as the slender vortical structures in the circumferential shear region are same in both cases.
The signs and strengths of the vortices are also accurately captured in the current approach. 
The circumferential shift of the lobes in the primary vortex region is due to different size of subspace used in each method, however the shapes of most unstable modes ($\omega_3, \omega_4$ resemble very close in terms of their circumferential positions. 

\subsection{Three-Dimensional Lid Driven Cavity (3DLDC)}
Motivated by the computational performance of our current approach, we now show its ability to solve a high degree of freedom system in a three-dimensional lid-driven cavity. 
The examination of linear stability of the full confined 3DLDC with classical methods, has been constrained until recently because of computational limitations. 
\citet{giannetti2009linear} performed linear stability analysis on this flow, although for a relatively coarse grid.
Their results suggest that the flow becomes unstable at $Re\approx 2000$. 
Subsequent efforts have examined the instability mechanism in more detail, including that of second Hopf bifurcation, using variants of time-stepper approach \citep{Gomez2014, Loiseau2016, Picella2018}. 
In the 3DLDC setup, all sides of the cube are stationary walls, except the lid which moves at a constant velocity.  
A cubical computational domain of $[0~~1]^3$ is considered, discretized with $121$ grid points in each direction. 
Simulations are performed at Reynolds numbers of $Re=2,100$, which is slightly larger than the critical Reynolds number for the incompressible case as noted in the experimental and numerical studies of \citet{liberzon2011experimental}. 
The non-dimensional time-step used for the simulation is $\Delta t = 2\times 10^{-4}$. 

\begin{figure}
\centering
\subfloat[Streamlines at mid-plane show the primary vortex, and two secondary vortices near the upstream and downstream walls at the bottom.]{\includegraphics[width=.45\textwidth]{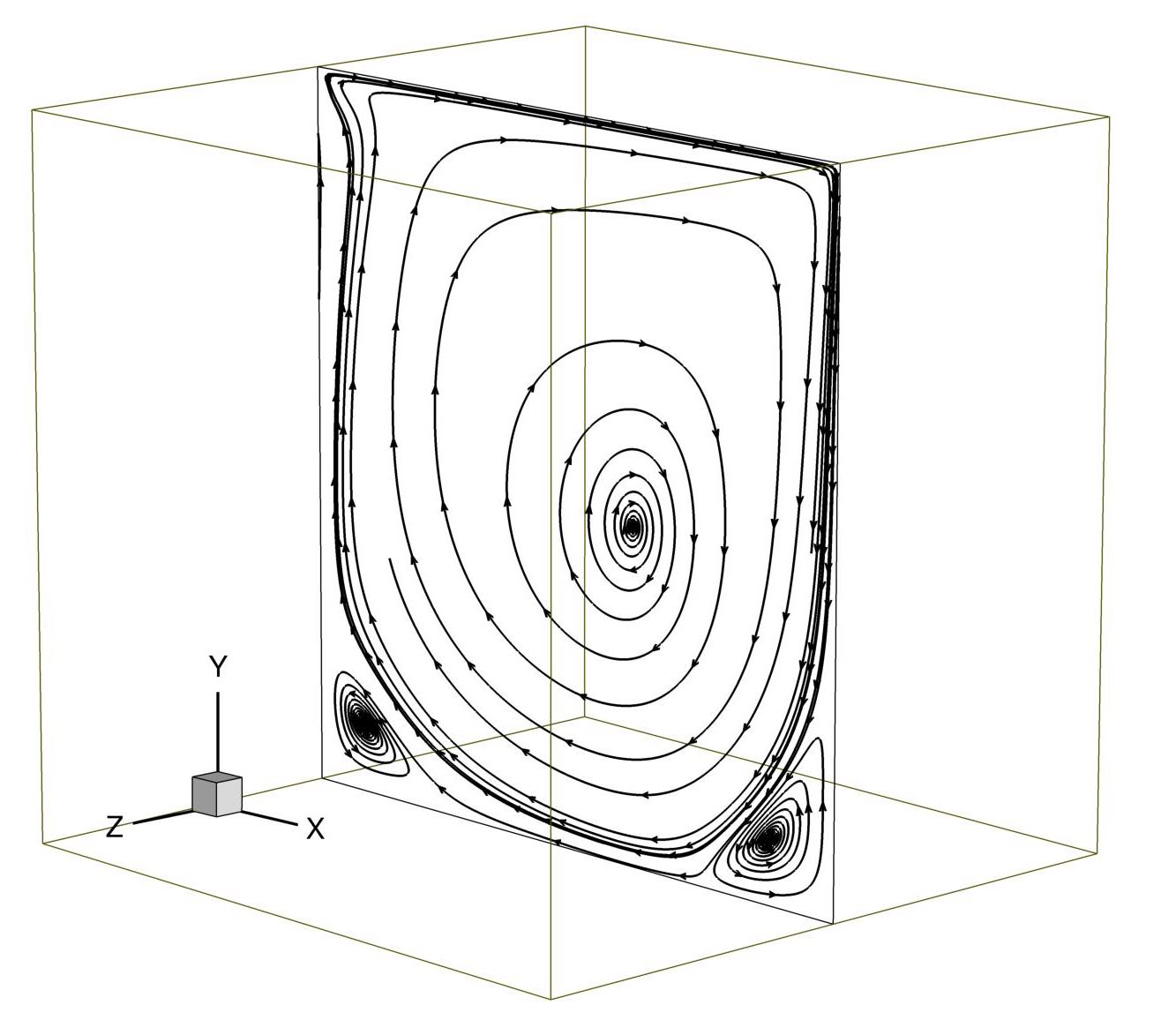}}\qquad
\subfloat[$u$ and $v$ profiles plotted at $(0.5,y,0.5)$ and $(x,0.5,0.5)$ respectively, and compared with \citet{giannetti2009linear} (circles)]{\includegraphics[width=.45\textwidth]{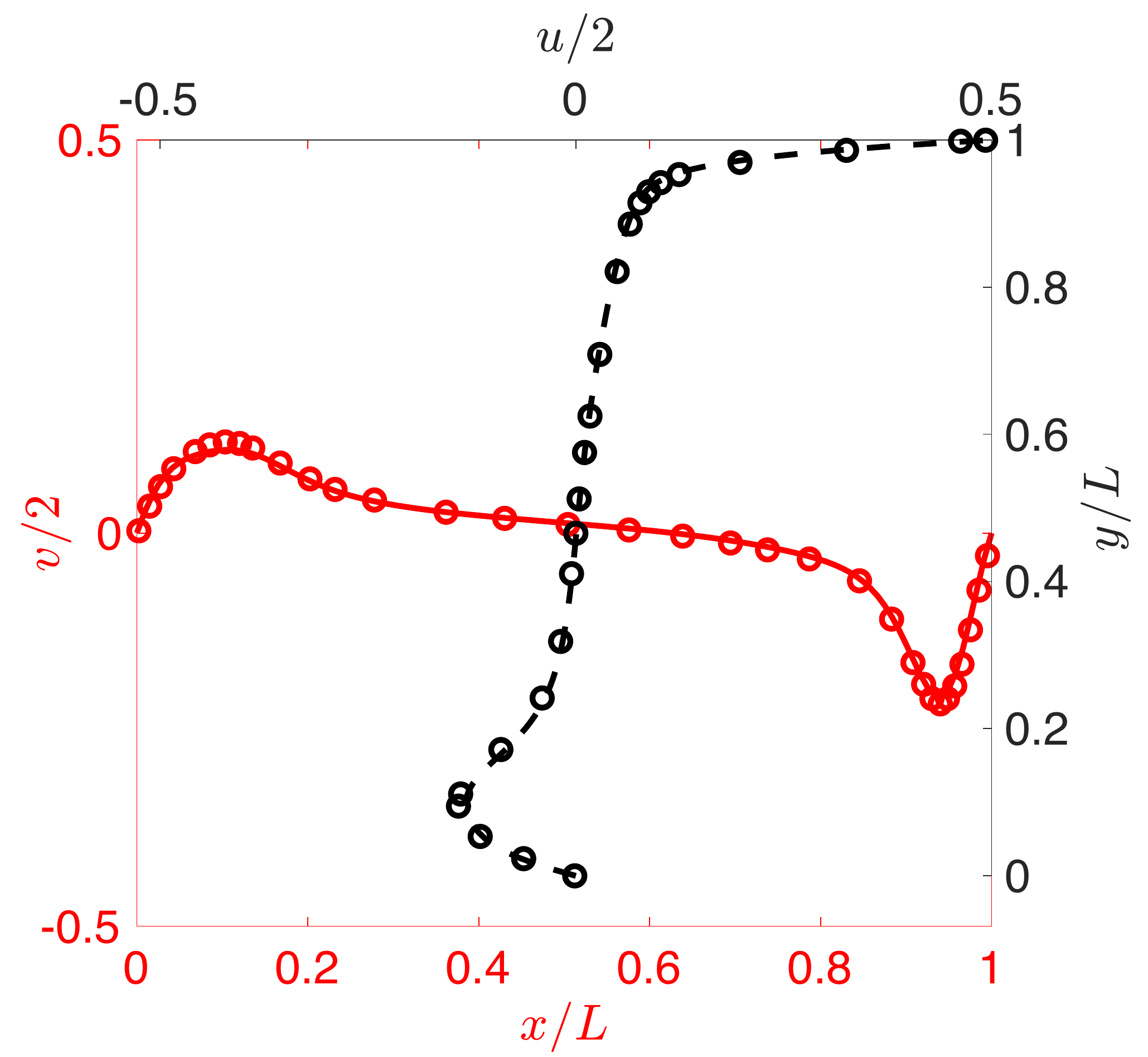}}
\caption{Basic state for three-dimensional LDC stability study.}
\label{fig:3dlc1}
\end{figure}

The basic state, shown in Fig. \ref{fig:3dlc1}, is obtained by long time-averaging instantaneous snapshots after a statistically stationary state is reached. 
The streamlines on the mid-plane in Fig.~\ref{fig:3dlc1}(a) indicate some similarities to the 2D counterpart, including a large central vortex and corner eddies. 
However, the stability properties are different as discussed further below.
Figure~\ref{fig:3dlc1}(b) shows a sample quantitative validation of the basic state with the numerical data presented in \citet{giannetti2009linear} at $Re=2000$, which was used for stability study. 
The $u$- and $v$-profiles are plotted along centerlines $(0.5,y,0.5)$ and $(x,0.5,0.5)$ respectively.
The current basic state compares very well with the reference data, and is now employed for stability analyses.

\begin{figure}
\centering
\subfloat[$t=0.04$] {\includegraphics[width=.33\textwidth]{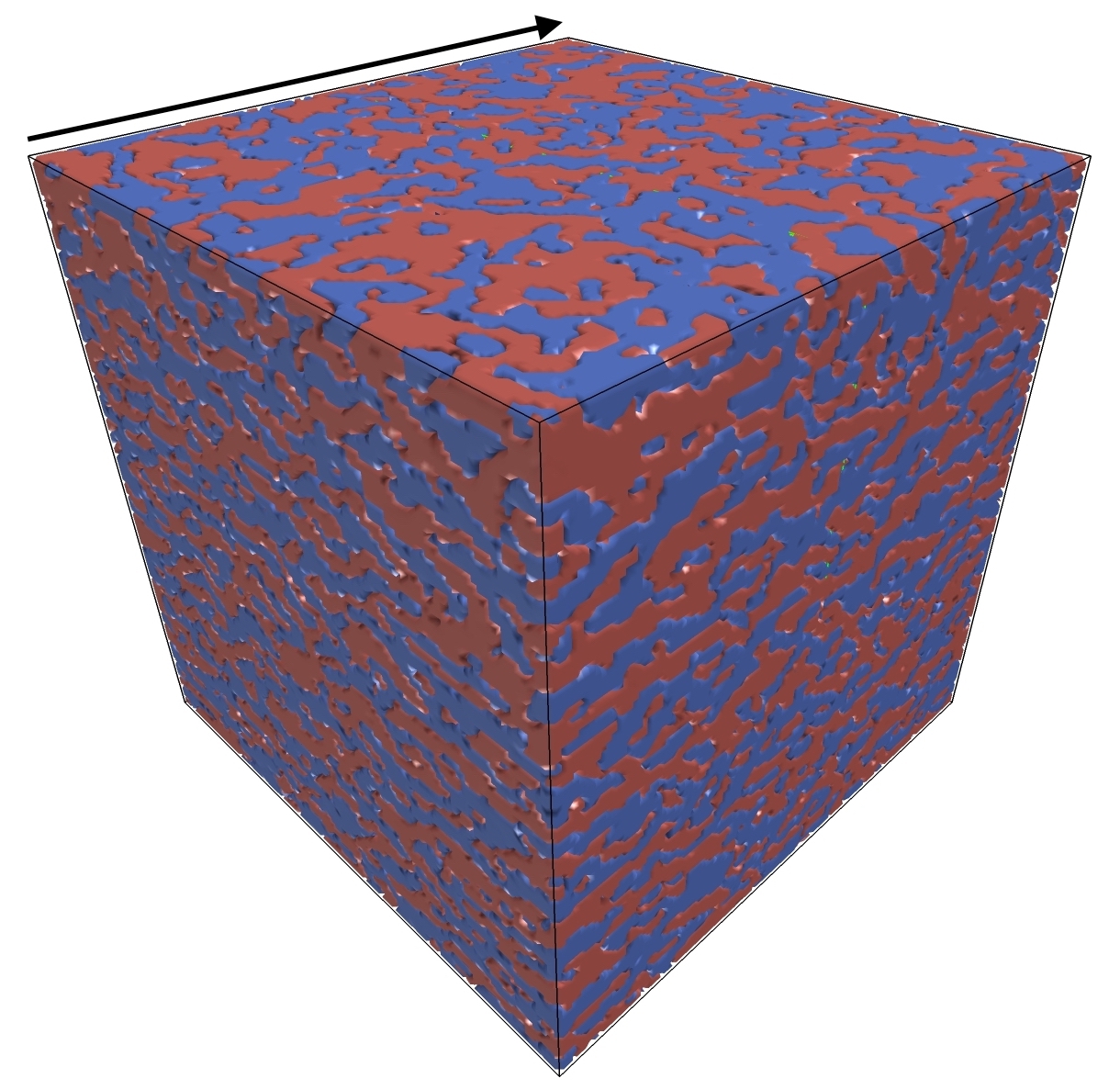}}
\subfloat[$t=6$] {\includegraphics[width=.33\textwidth]{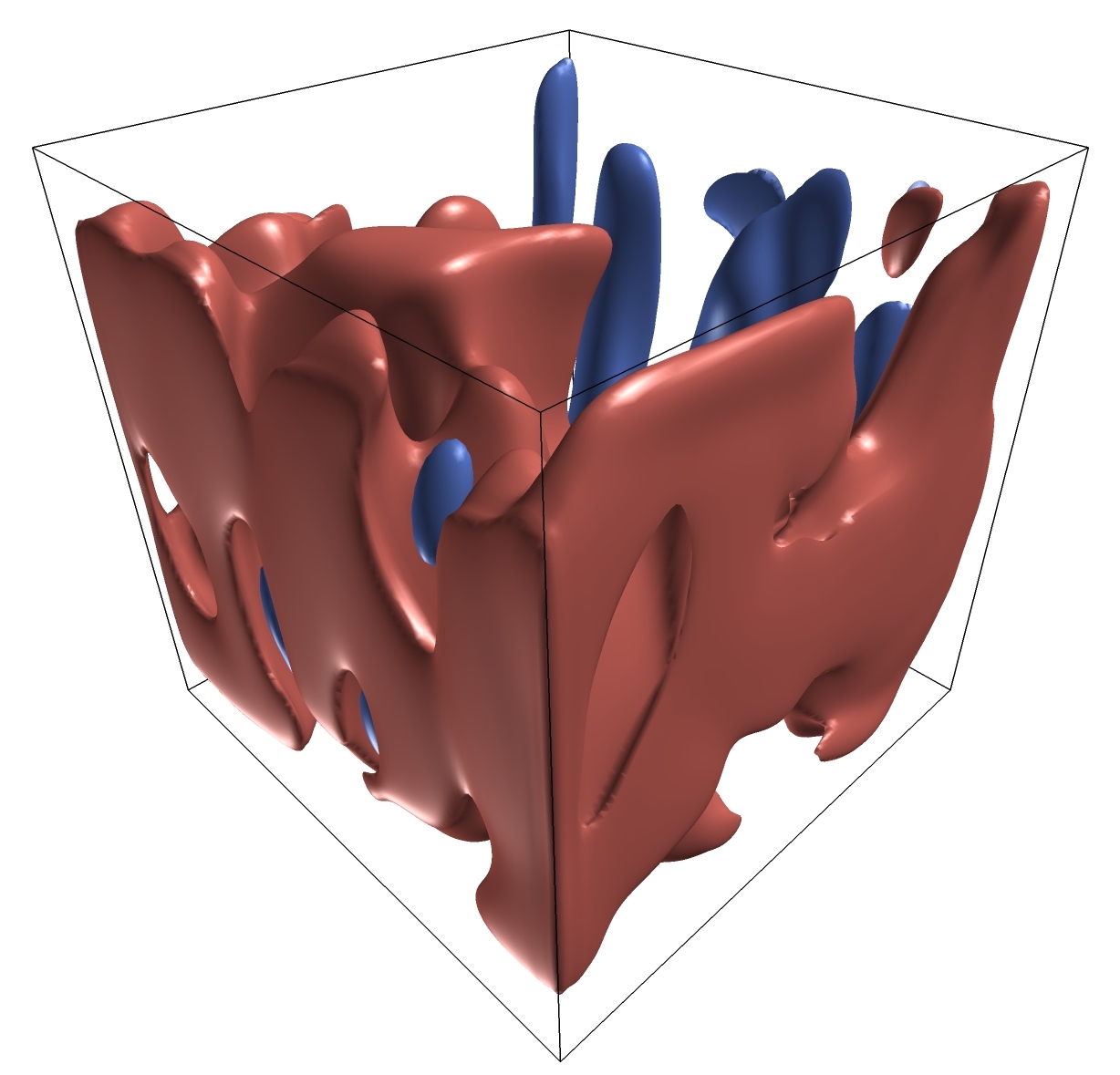}}
\subfloat[$t=60$] {\includegraphics[width=.33\textwidth]{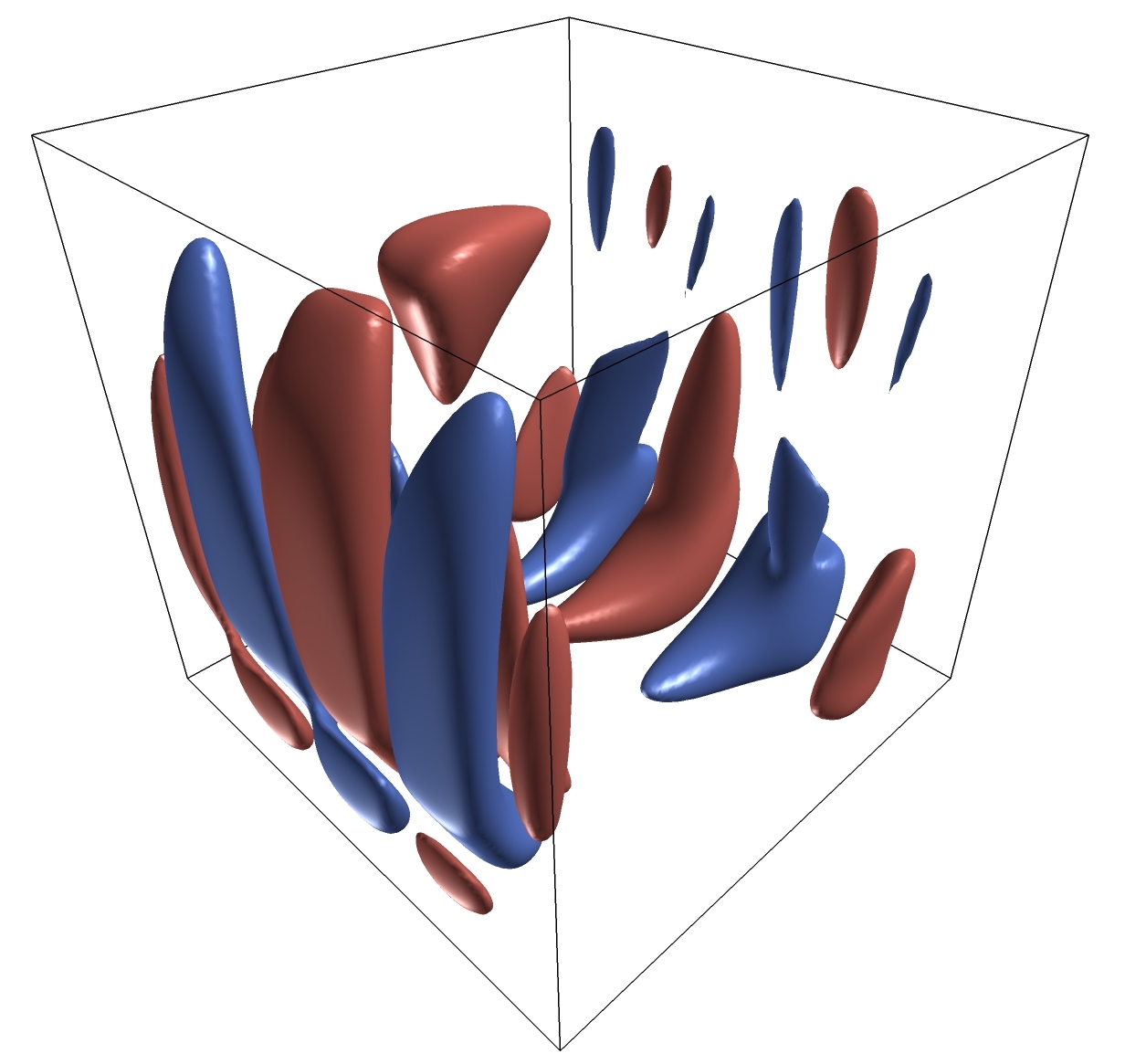}}
\caption{Evolution of perturbations using NS-MFP shown through $v^\prime$ iso-surfaces (3DLDC).}
\label{fig:3dlc2}
\end{figure}
The linear evolution of the perturbation using the basic state and an initial volumetric random field is shown using $v'$-velocity iso-surfaces in Fig. \ref{fig:3dlc2}.  
After a flow time of $t=16$, the characteristic structures completely fill the domain. 
The perturbations evolve into elongated banana-shaped structures, resembling Taylor-G\"{o}rtler vortices, near the walls. 

\begin{figure}
\centering
\includegraphics[width=0.8\textwidth]{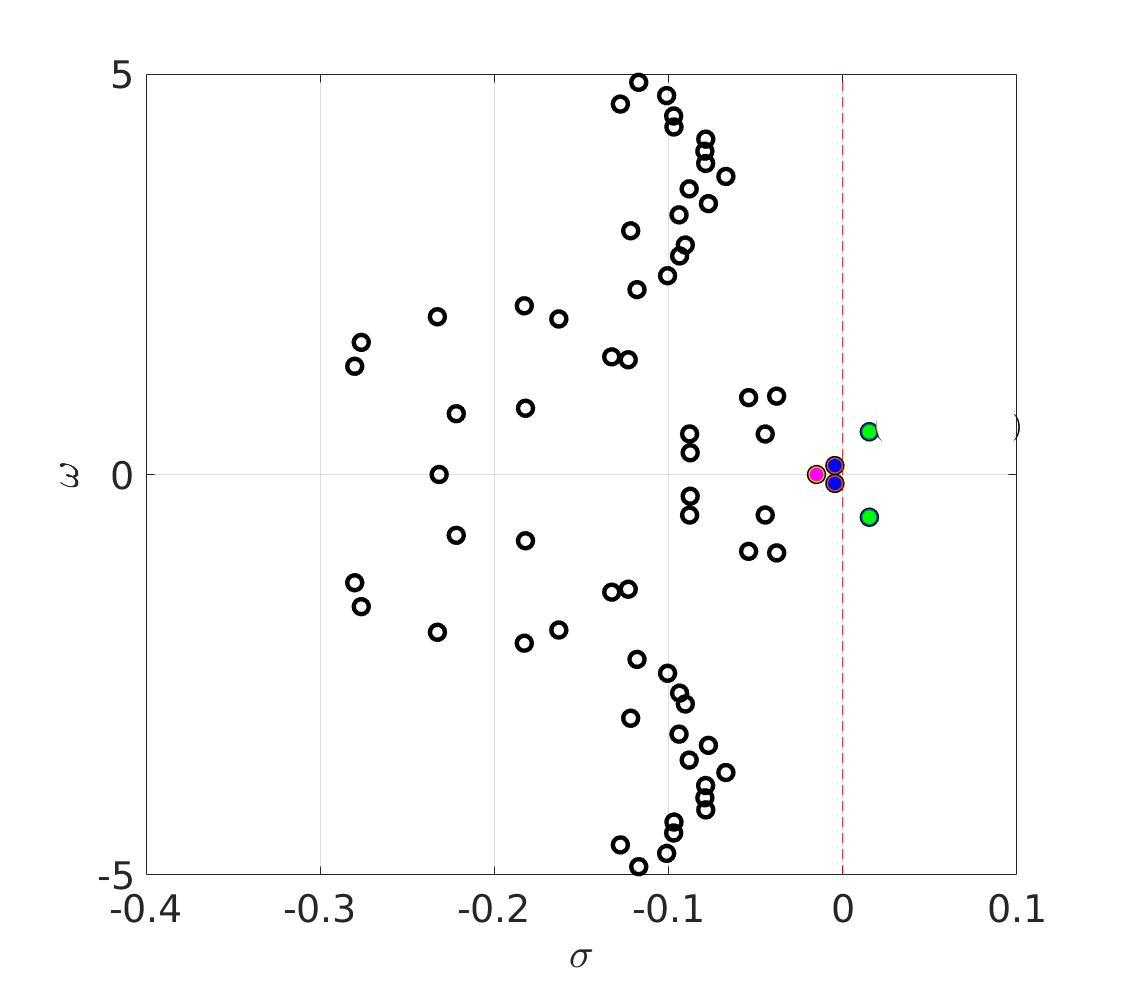}
\caption{Eigenspectrum for 3DLDC stability study. Green and blue circles indicate primary and secondary stability modes. Magenta circle is the least stable stationary mode.}
\label{fig:3dlc_eig}
\end{figure}
\begin{figure}
\centering
\subfloat[$\omega_c\simeq0.55$]{\includegraphics[width=.33\textwidth]{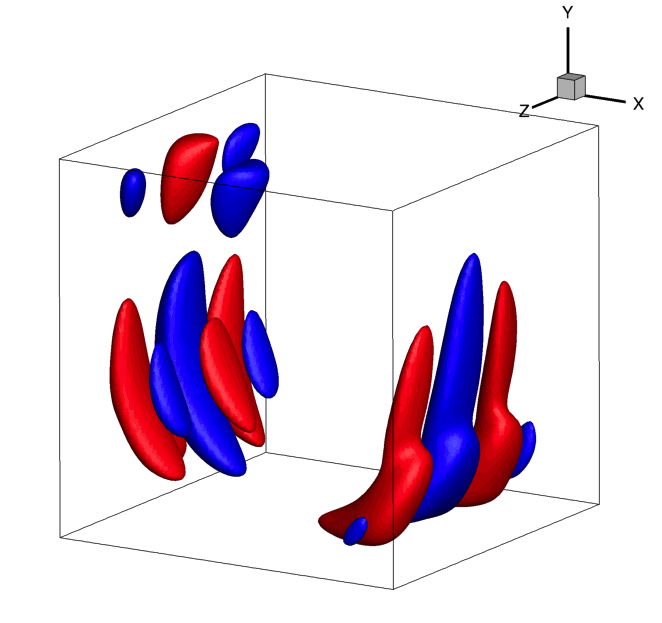}}
\subfloat[$\omega_c\simeq0.12$]{\includegraphics[width=.33\textwidth]{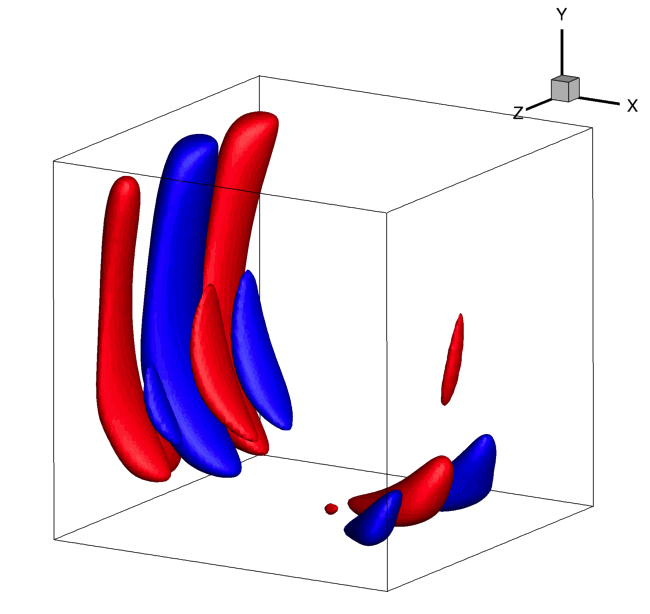}}
\subfloat[$\omega_c=0$]{\includegraphics[width=.33\textwidth]{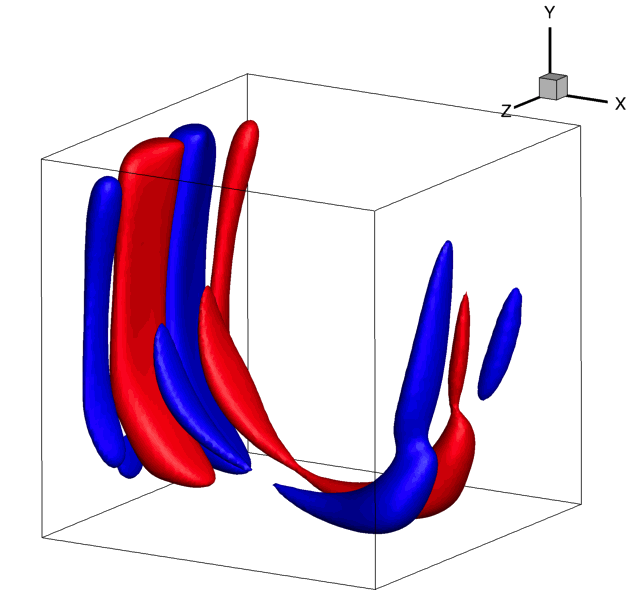}}
\caption{Dominant oscillatory and non-oscillatory instablity modes in 3DLDC. Iso-surfaces of $v'$ are shown.}
\label{fig:Re2100_modes}
\end{figure}
The eigenspectrum obtained with DMD using 1000 snapshots with all three velocity components is shown in Fig. \ref{fig:3dlc_eig}. 
The most unstable mode is located at a circular frequency of $\omega_c = 0.55$, which is close to the reported value $\omega_c=0.575$\citep{Feldman2010} for the incompressible 3DLDC. 
The slight dip in the frequency is also postulated to be the effect of compressibility as mentioned earlier. 
Besides this most unstable mode, two secondary stability modes, shown in blue and magenta colors in Fig. \ref{fig:3dlc_eig}, are observed.  
Of these two modes, the one at $\omega_c \simeq 0.12$ is oscillatory, while other is  stationary. 

The importance of these stability modes has been discussed in  studies on the different competing mechanisms that exist in this flowfield. 
The frequency of the oscillatory mode agrees with that observed by \citet{Loiseau2016}, who detect the signature of this instability in the form of chaotic intermittent events in long time simulations. 
The oscillatory mode, like the primary instability mode, is symmetric.
However it gains prominence over the primary mode when the mirror symmetry is broken post-bifurcation due to three-dimensional effects. 
On the other hand, the stationary mode results primarily due to symmetry-breaking and was earlier speculated to be primary mechanism for bifurcation in 3DLDC \citep{giannetti2009linear}.
\citet{Feldman2010} note however that symmetry-breaking gains prominence only after bifurcation. 
Though a detailed discussion of these instabilities is out of scope for this paper, the results highlight the capability of the current approach in extracting these leading instabilities without input parameter guesses. 

We close the discussion by showing the spatial structures of these modes in Fig.~\ref{fig:Re2100_modes}. 
The most unstable mode displays counter-rotating vortices at the upstream wall as well as streaks on the opposite side walls.  
The structure of this mode is symmetric and recovers that of the leading unstable mode found in \citet{Gomez2014} and \citet{Feldman2010} for incompressible flows. 
On the other hand, the secondary oscillatory mode shown in Fig.~\ref{fig:Re2100_modes}(b), is dominated by long streaks on the upstream wall, while the  counter-rotating vortices near the top wall are absent. 
Like the primary mode, this low-frequency mode is also symmetric, indicating that it may be the one responsible for chaotic intermittent events observed in \citet{Loiseau2016}. 
The stationary mode, depicted in  Fig.~\ref{fig:Re2100_modes}(c), however is anti-symmetric, which confirms the post-bifurcation  breakup of symmetry of the basic state.

\subsection{Two-Dimensional Cylinder (CYL2D)}
\begin{figure}
\centering
\includegraphics[width=.8\textwidth, trim={1cm 2cm 2cm 4cm},clip]{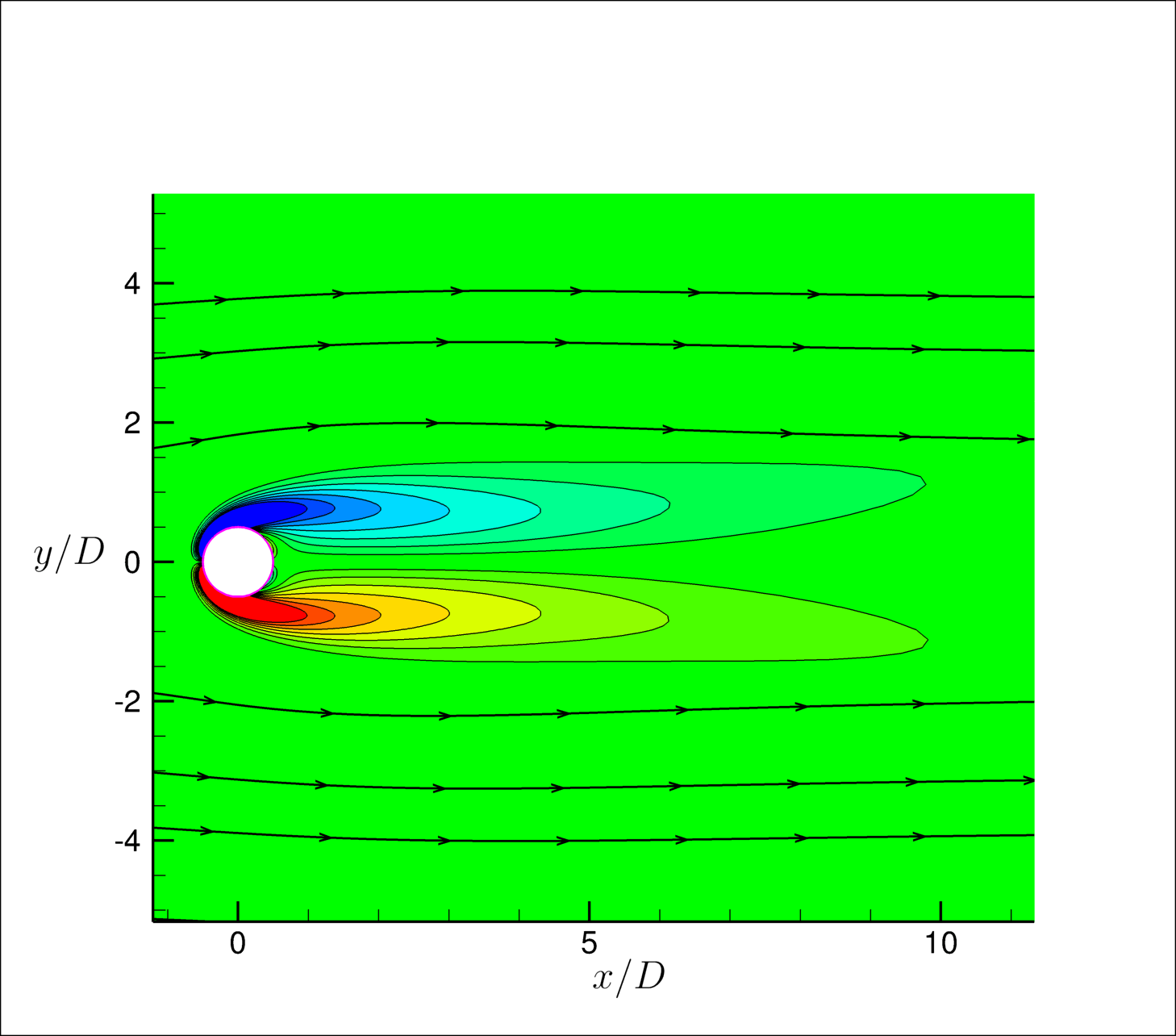}
\caption{Basic state for cylinder flow used for stability study ($Re=50$).}
\label{fig:cyl1}
\end{figure}
\begin{figure}
\centering
\subfloat[$t=0.05$]{\includegraphics[width=.7\textwidth]{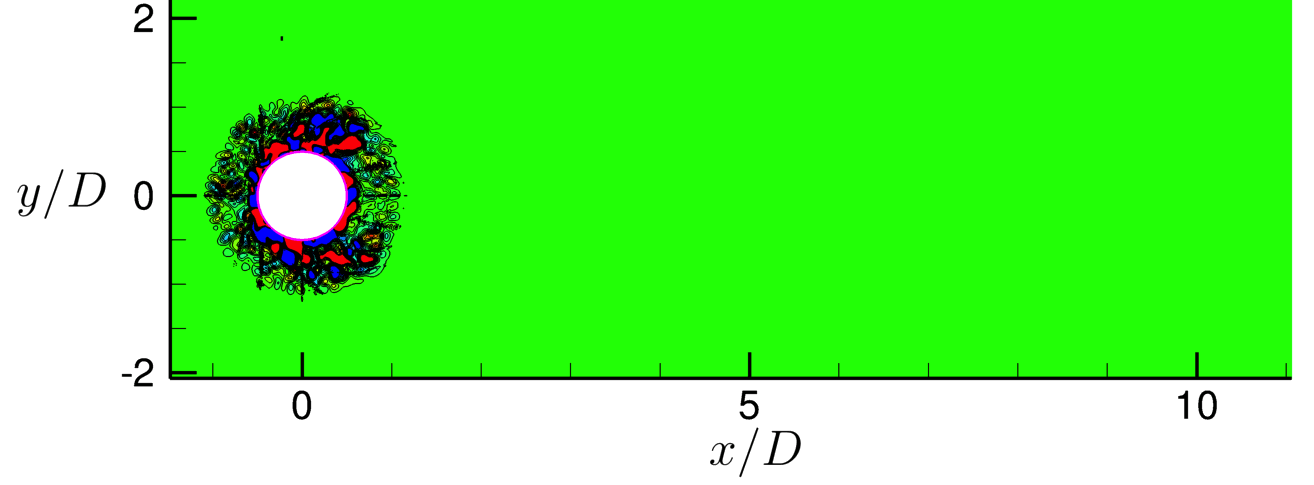}}\\
\subfloat[$t=2$]{\includegraphics[width=.7\textwidth]{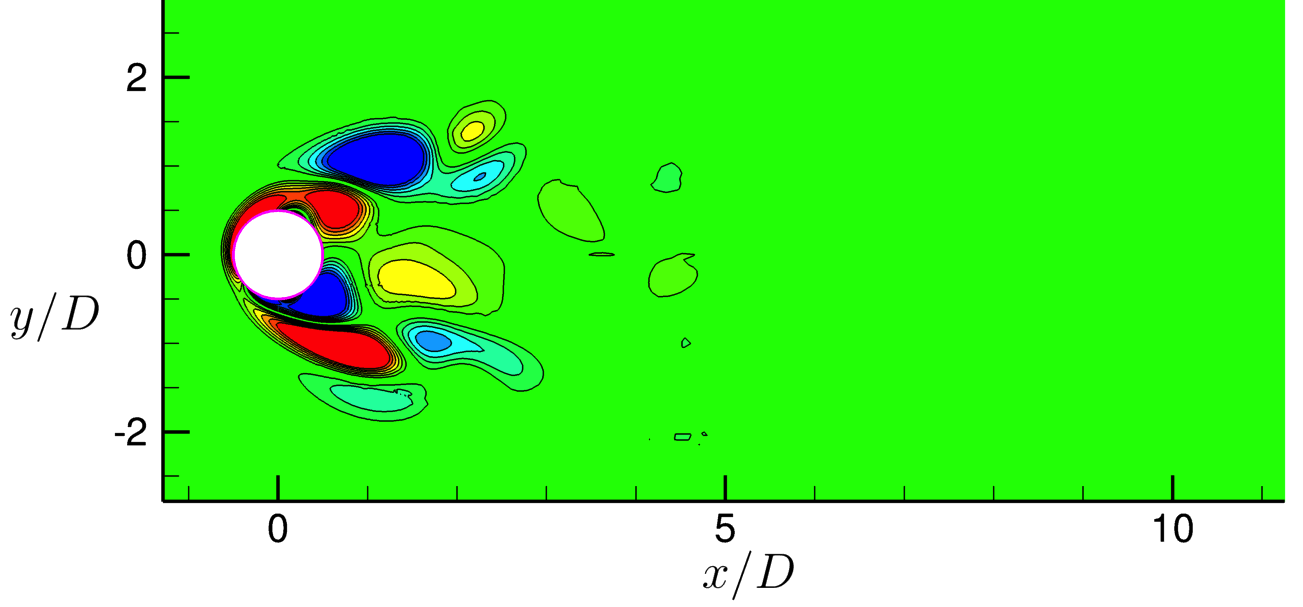}}\\
\subfloat[$t=10$]{\includegraphics[width=.7\textwidth]{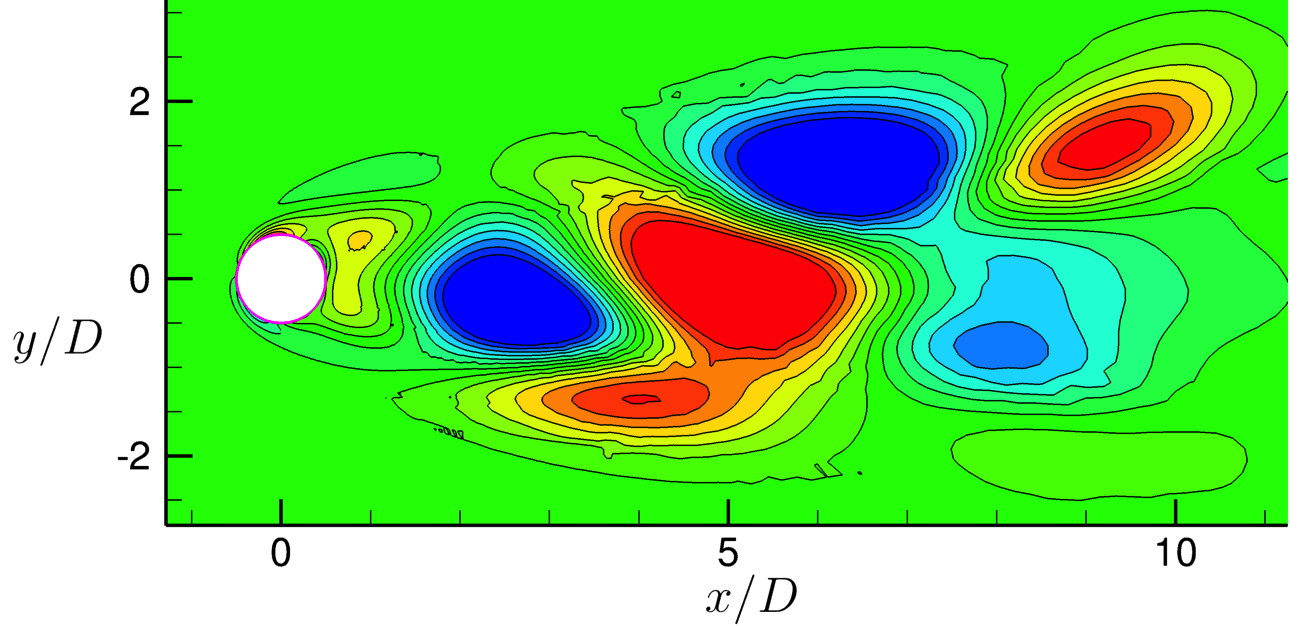}}
\caption{Evolution of perturbation snapshots from a localized forcing near the cylinder body.}
\label{fig:cyl_inst}
\end{figure}
As an example of an external flow, we consider the flow past a circular cylinder, which is representative of many engineering applications. 
The flow exhibits rich dynamics and its evolution may be characterized by the free-stream Reynolds number $Re$ based on cylinder diameter $D$. 
For example, when $Re<47$, a steady and symmetric closed wake is formed in the recirculation region behind the cylinder \citep{roshko1993perspectives}.
As the Reynolds number is increased, the flow becomes unsteady  and instabilities appear, eventually leading to the periodic shedding of laminar vortices (von K\'arm\'an vortex shedding), resulting in increased mean drag force and ultimately turbulence.

The critical Reynolds number, $Re_{cr}$, where the two-dimensional steady state changes to a two-dimensional periodic limit cycle, has been explored in many studies \cite{roshko1954development, kumar2006prediction, sipp2007global, bagheri2013koopman}.  
In the present exercise, we use our approach to obtain the stability modes at two Reynolds numbers based on diameter, $D$, of $Re=40$ and $50$, which lie on the either side of $Re_{cr}$, based on reported values for incompressible flows.

The basic state is obtained on a circular domain extending $1,000D$ around the cylinder.
Based on a grid resolution study, the discretization uses $500$ and $870$ points in the azimuthal and radial directions respectively.
Uniform angular spacing is used in the azimuthal direction while a stretched mesh discretizes the radial direction with the first grid point located at $0.005D$ from the cylinder surface. 
At the farfield, a characteristic-based boundary condition is used to ensure smooth outflow.   
Elements of the nature of the basic state are shown for $Re=50$ in Fig.~\ref{fig:cyl1} with vorticity contours. Since the $Re$ considered here is near the critical Reynolds number, no visible asymmetry is observed in the wake region. 

The procedure to generate snapshots is the same as discussed above for LDC flows with the exception that instead of applying pulsed random forcing to the entire volume, here we apply it only in the circular region of diameter $2D$ around the cylinder surface. 
The presence of vortex shedding in the cylinder flow, as well as the dominance of the near body dynamics ensures that localized forcing is sufficient.
Representative perturbation snapshots are shown in terms of vorticity in Fig. \ref{fig:cyl_inst}. 
As time progresses, the initial localized disturbance fills the entire computational domain.
Also, structures in the wake region of the flow become more organized. 
When these snapshots are subjected to DMD, detailed information about the stability modes and their frequency properties are obtained in a straightforward manner. 

\begin{figure}
\centering
\subfloat[$Re=40$]{\includegraphics[width=.47\textwidth]{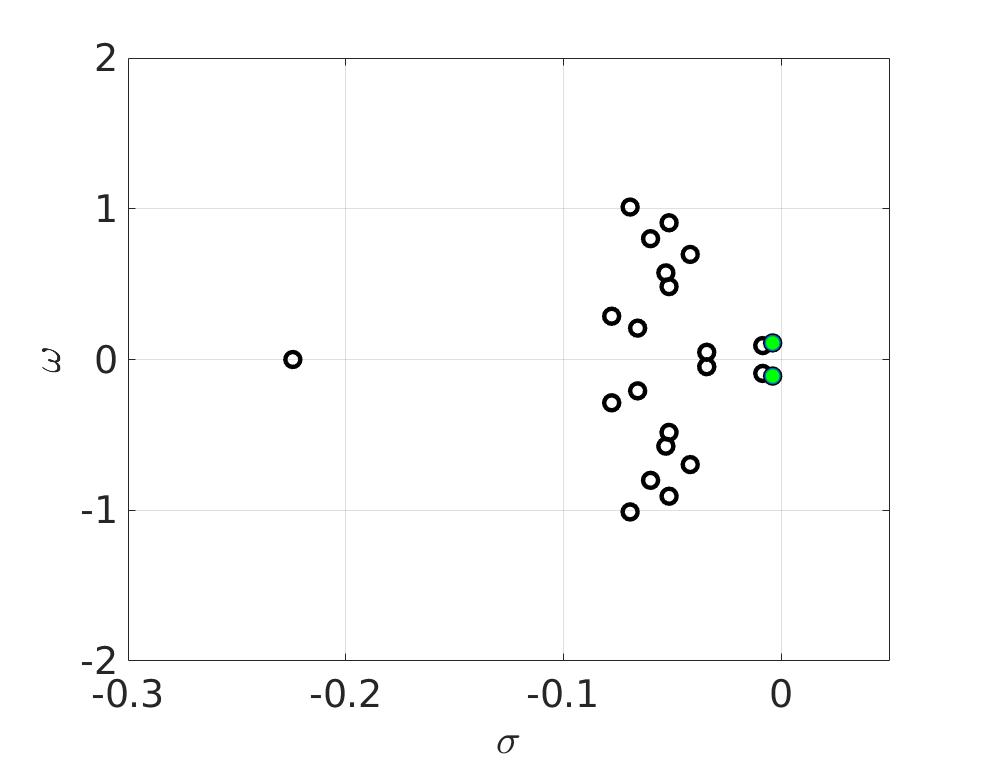}}
\subfloat[$Re=50$]{\includegraphics[width=.48\textwidth]{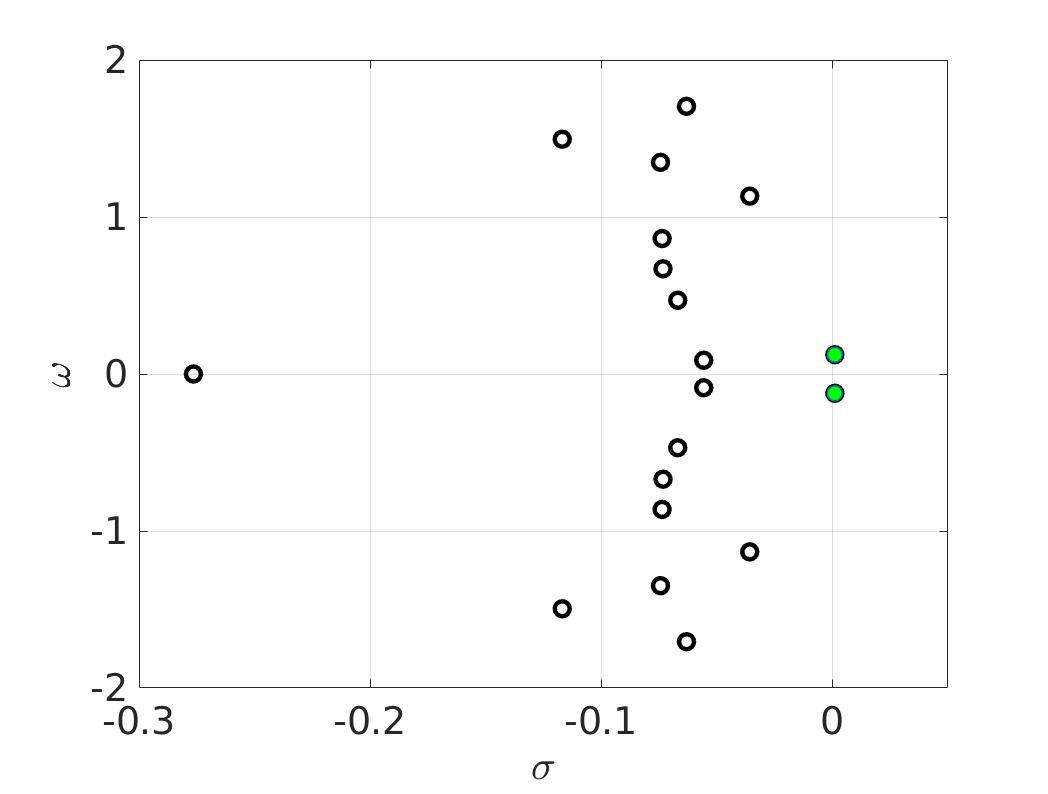}}
\caption{Eigenspectra in stability studies of flow past a cylinder. The least stable mode at $Re=40$ in (a) becomes unstable at $Re=50$ in (b)}
\label{fig:cyl_eig}
\end{figure}
For the analysis, it is sufficient to use a domain with only $500$ grid points in the radial direction.
The eigenspectra thus obtained for both Reynolds number flows are shown in Fig. \ref{fig:cyl_eig}.  
The frequency of the least stable mode for $Re=40$ is $\omega~\simeq~0.12$, agrees with earlier studies using different stability approaches (see \cite{kumar2006prediction} for a list). 
When the $Re$ is increased to 50, this decaying mode crosses the vertical axis and manifests as a growing mode, as shown in Fig.  \ref{fig:cyl_eig}(b). 
This indicates that at this Reynolds number, bifurcation has already occurred, as is consistent with the observations in the literature, where the  $Re_{cr}$ is found to be in the range of $46$ to $48$ in incompressible studies.  

\begin{figure}
\centering
\includegraphics[width=.9\textwidth]{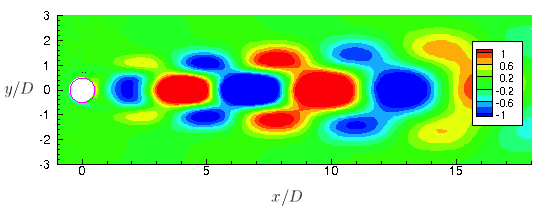}
\caption{Most unstable eigenmode for flow past a cylinder ($Re=50$).  Contours of vorticity perturbation are plotted.}
\label{fig:cyl3}
\end{figure}
Figure \ref{fig:cyl3} shows the vorticity field for the most unstable mode at $Re=50$. 
The alternating vorticity field in the wake region of the cylinder is typical of the flow at this Reynolds number. 
The shape is also strikingly similar to the published results of \citet{kumar2006prediction}. 
The form of the least stable mode for $Re=40$ is qualitatively very similar to  Fig. \ref{fig:cyl3}, hence is not shown for brevity.

\subsection{NACA0015 Airfoil at Stalled Conditions (NACA2D)}\label{sec:naca0015}
As a final example of the ease with which primary stability modes can be obtained for more complex situations, we consider linear stability of a stalled symmetric NACA airfoil (0015).
For the stability analysis, the Reynolds number considered is relatively low ($Re=200$), but the angle of incidence ($\alpha=18^{\circ}$) is sufficiently high so that flow on the suction side is separated near the trailing edge. 
The main interest in this flow is to identify the primary linear mechanism for unsteadiness, as there are two competing dynamics:  amplification of Kelvin-Helmholtz wake mode, and self-excitation of laminar separation bubble appearing as a stationary mode \cite{theofilis2000origins}. 
Though a detailed investigation into the latter requires a three-dimensional stability study, for the current validation purpose we restrict ourselves to two-dimensions in which the wake mode is found to be independent of spanwise wavenumber parameters \cite{gioria2015global}.

Results from both matrix-forming as well as time-stepper approaches \cite{kitsios2009biglobal, rodriguez2011birth, gioria2015global, he2017linear} are available for comparison, albeit in an incompressible framework.  
While earlier studies \cite{kitsios2009biglobal, rodriguez2011birth} have used a BiGlobal code with conformally mapped curvilinear co-ordinates to obtain eigenmodes,  subsequent studies \cite{gioria2015global, he2017linear} use open-source codes, including Nektar++ and Nek5000 for the time-stepper approach, and FreeFEM++ for matrix-forming. 
In these studies, the Arnoldi subspace iteration is used for the relevant eigenspectrum computation, with the goal of identifying the smallest modulus modes.  
However, an interesting point concerns the issue of convergence. 
For example, earlier studies found a growing stationary mode (in a three-dimensional setup), which dominates  the decaying wake mode. 
Detailed analysis performed later clarified the observations \citep{he2017linear} . 
The problem highlights the importance of the \textit{a priori} known shift value in Arnoldi-based techniques.
The elimination of spurious eigenvalues, discussed for this flow by \citet{kitsios2010recovery}, is also a hurdle in broadening the scope of stability studies for non-trivial flows.
 
\begin{figure}
\centering
\includegraphics[width=.9\textwidth]{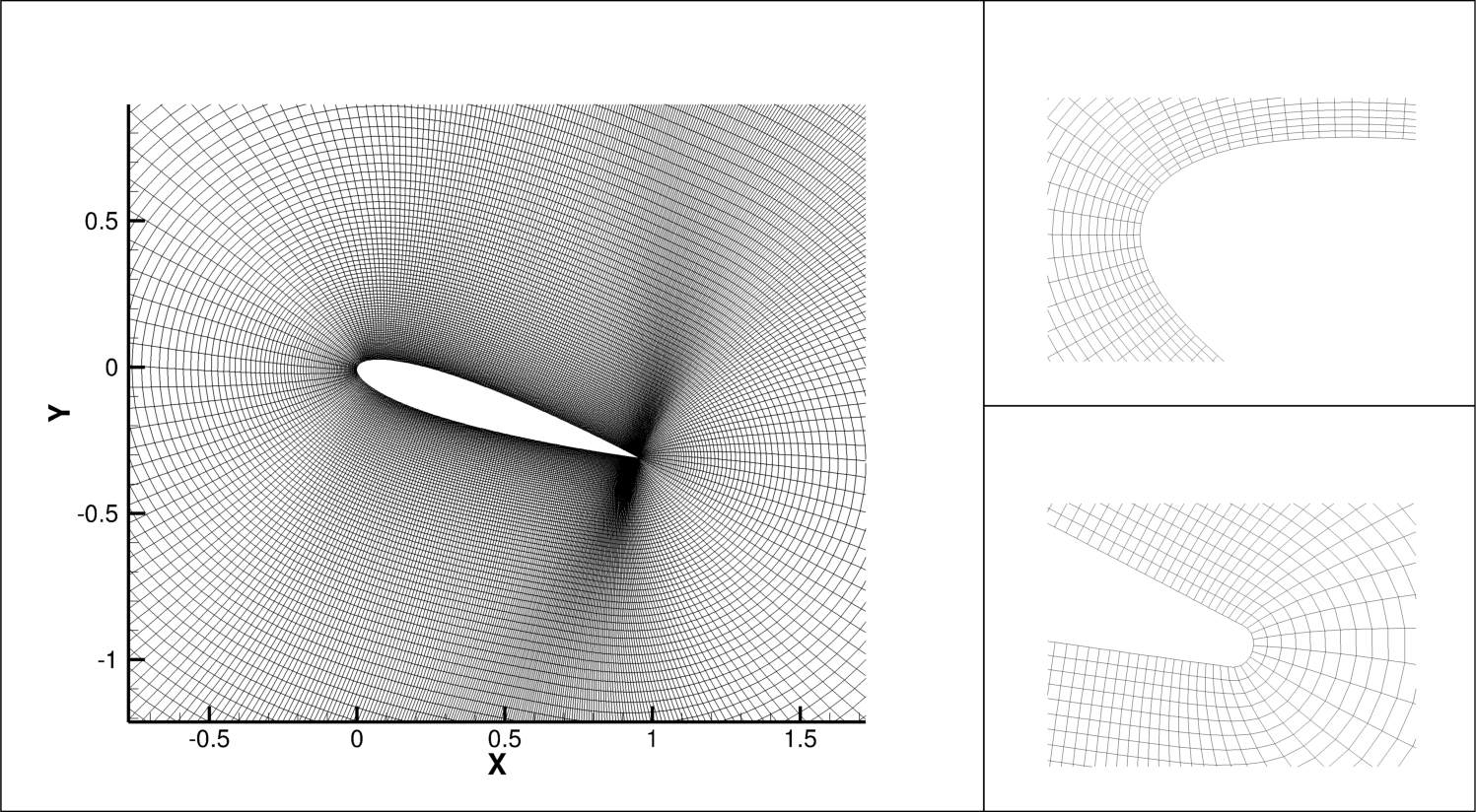}
\caption{Curvilinear grid used for NACA0015 base flow simulation and stability study. Every alternate grid point is shown. Zoomed-in region near the leading and trailing edges are shown on the right. }
\label{fig:naca_grid}
\end{figure}

We perform stability analysis of this flow at essentially incompressible ($M=0.1$) and higher ($M=0.5$) Mach numbers.
The former facilitates validation with the results of  \citet{he2017linear}, while the latter allows an estimation of the features of the method in the presence of compressibility.
An $O$-type structured grid is used around the airfoil (Fig. \ref{fig:naca_grid}), and the equations are solved in the curvilinear coordinate system.  
Unlike the sharp trailing edge used in the reference study, we use a more realistic circular trailing edge; \citet{he2015effect} note that such modifications have no appreciable effects on unsteadiness in the two-dimensional flow field. 
 
\begin{figure}
\centering
\subfloat[M=0.1] {\includegraphics[width=.49\textwidth]{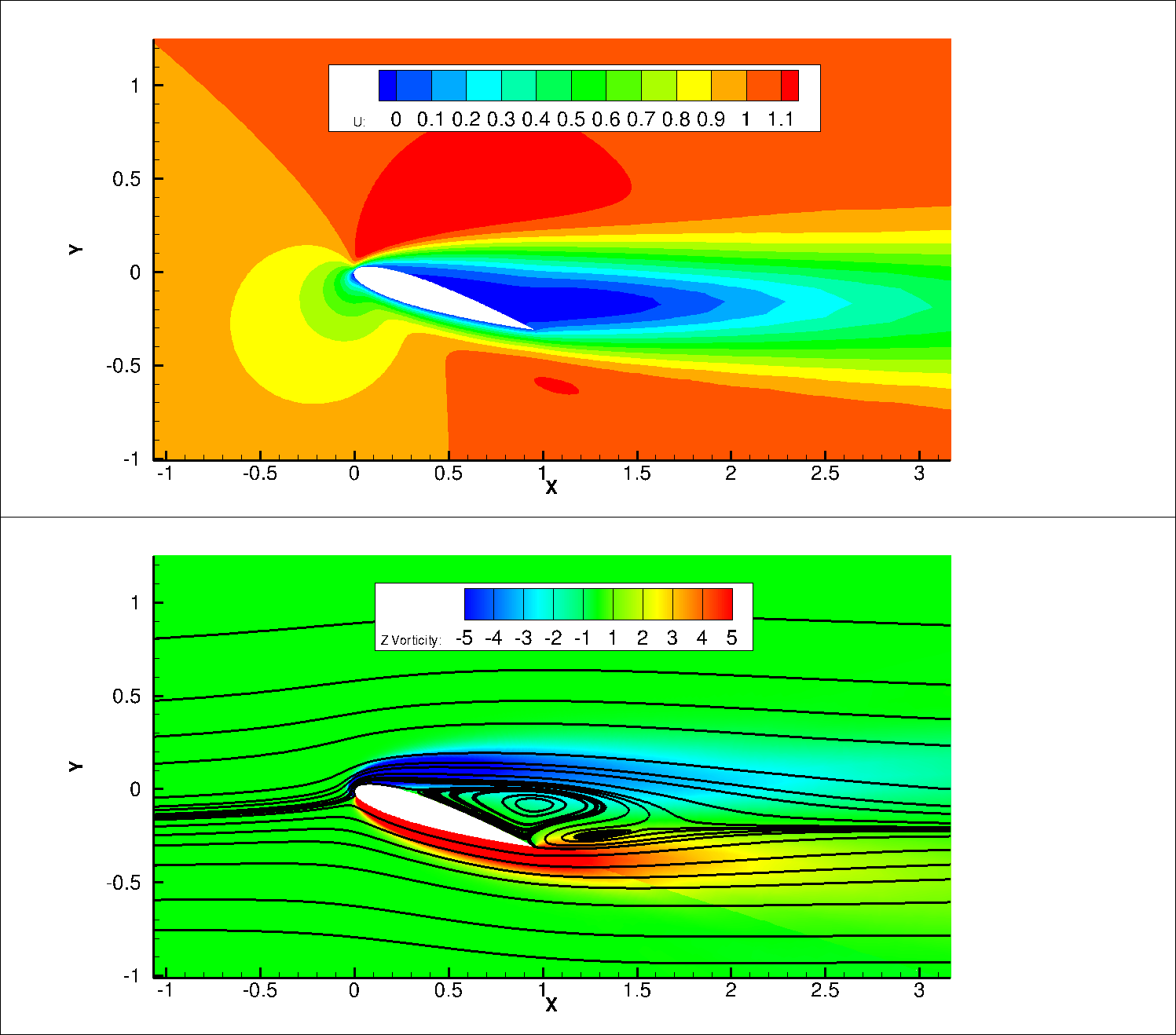}}
\subfloat[M=0.5] {\includegraphics[width=.49\textwidth]{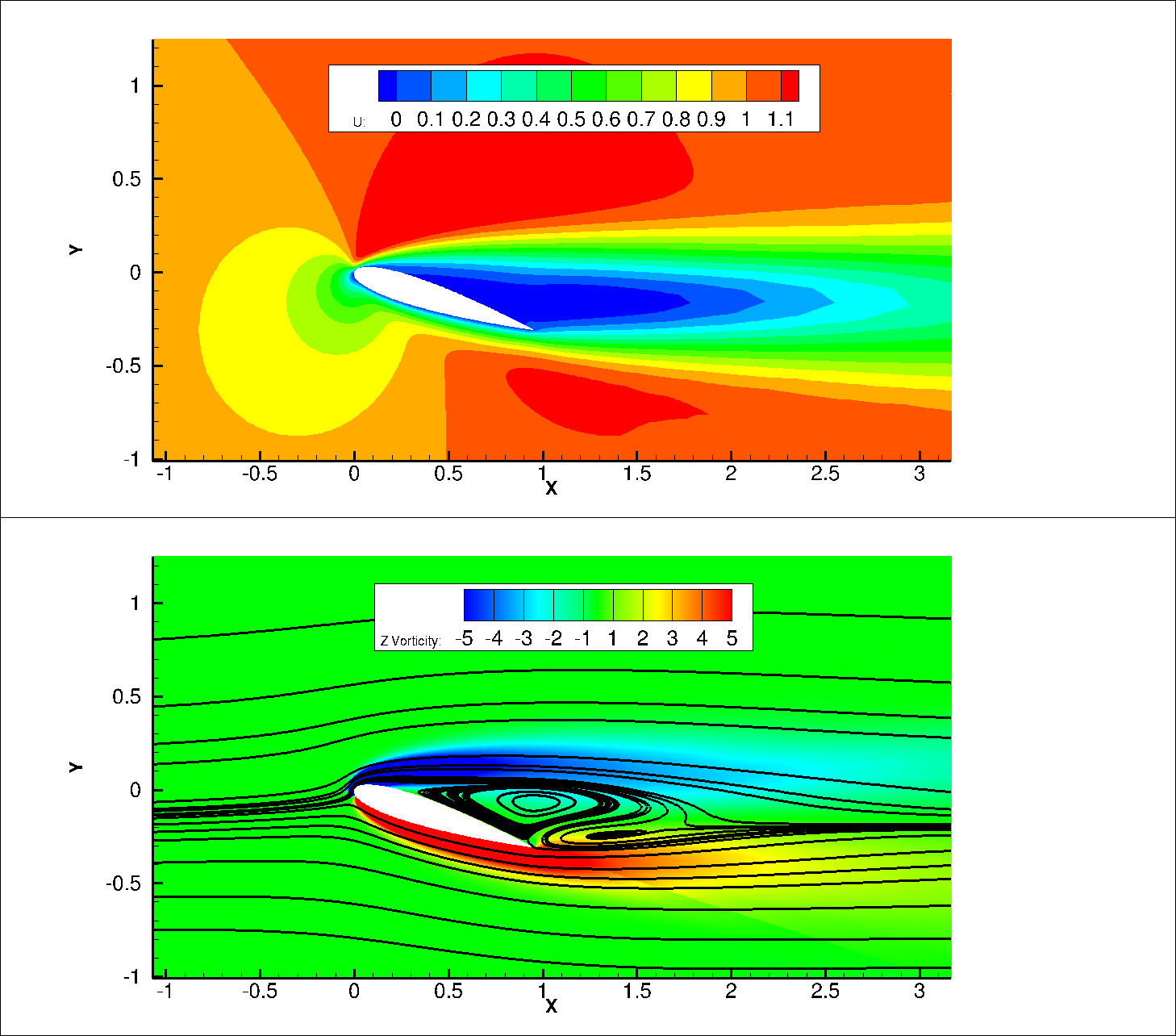}}
\caption{Basic state for flow past NACA0015 stability studies. Top: Contours of $x$-velocity. Bottom: Streamlines imposed on $z$-vorticity contours. }
\label{fig:naca_mean}
\end{figure}

Before showing the results for linear stability, we present the basic states for both Mach number cases in Fig.~\ref{fig:naca_mean}. 
Both basic states are steady and, as expected, show similar flow fields.
However, close examination reveals a slightly more elongated wake region in the high Mach number case. 
The separated flow fields obtained at such high incidence angles are clearly evident in the bottom panel.

\begin{figure}
\centering
\subfloat[M=0.1] {\includegraphics[width=.48\textwidth]{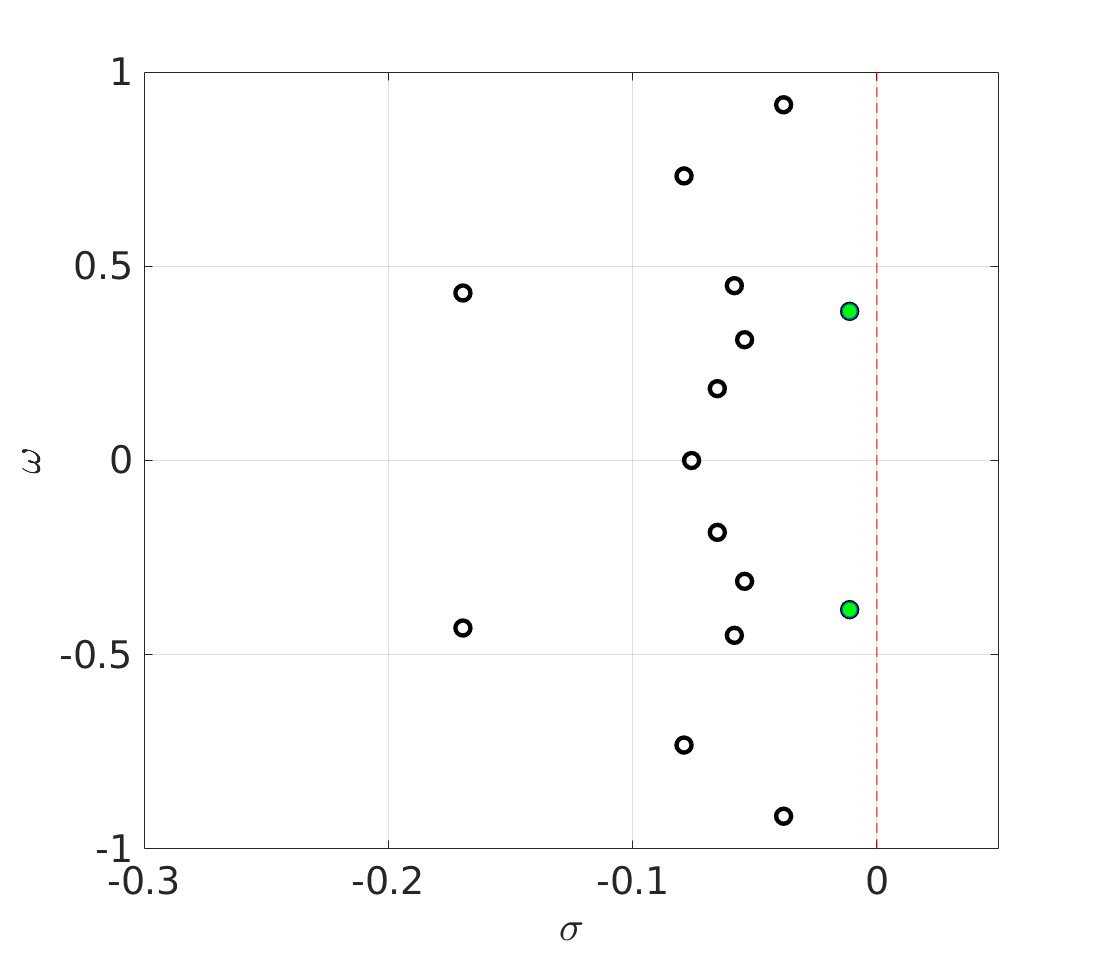}}
\subfloat[M=0.5] {\includegraphics[width=.48\textwidth]{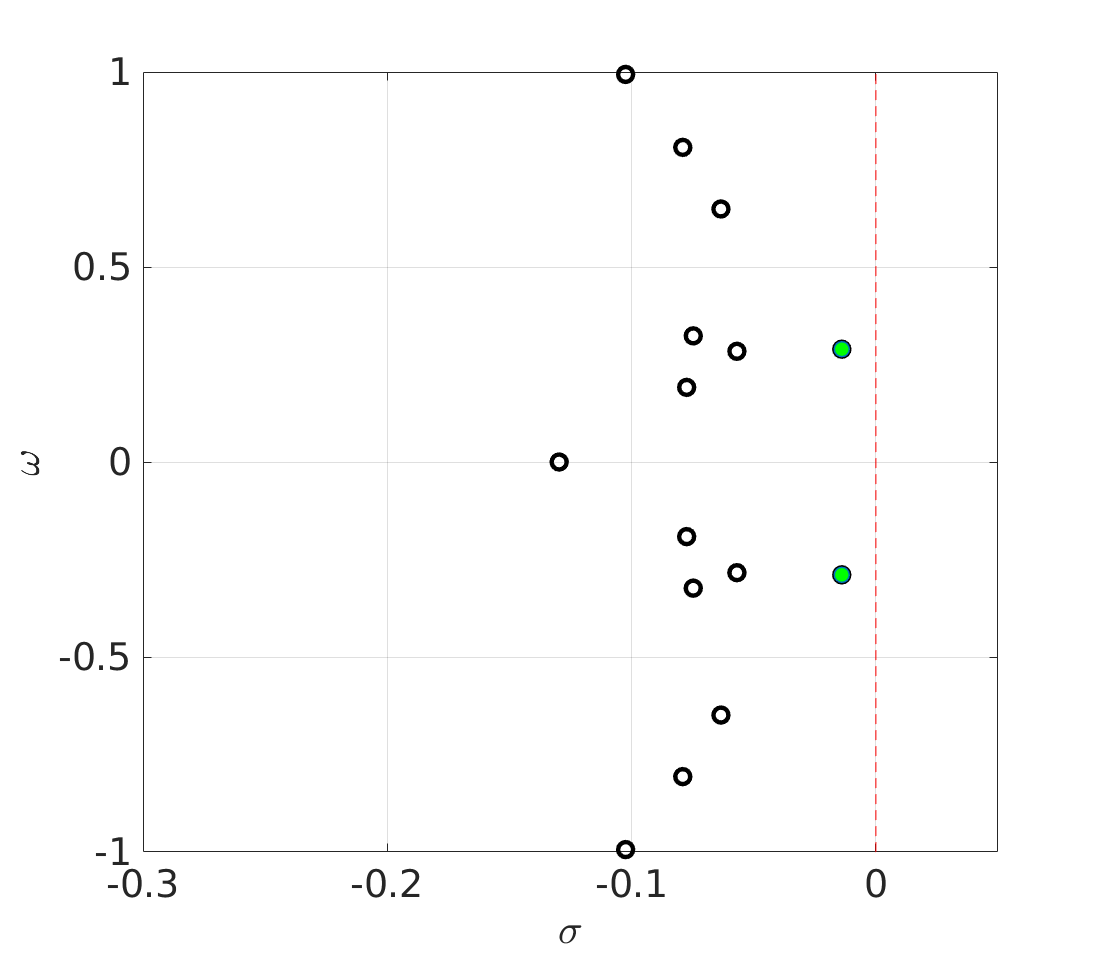}}
\caption{Eigenspectra in stability studies of a stalled NACA0015 airfoil flow}
\label{fig:naca_eig}
\end{figure}

Figure~\ref{fig:naca_eig} displays the eigenspectrum for both cases using the current approach. 
\begin{figure}
\centering
\subfloat[M=0.1] {\includegraphics[width=.48\textwidth]{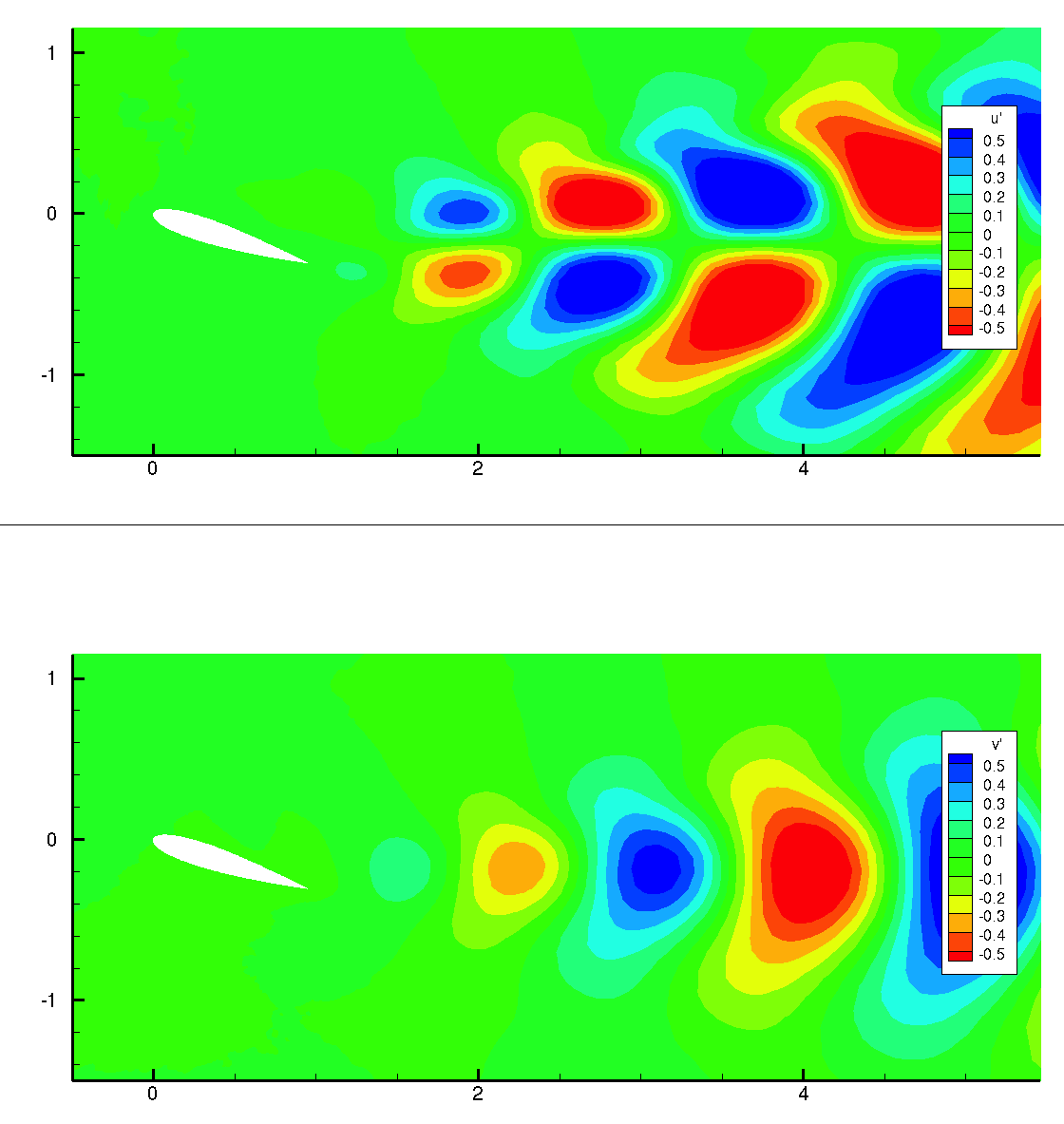}}
\subfloat[M=0.5] {\includegraphics[width=.48\textwidth]{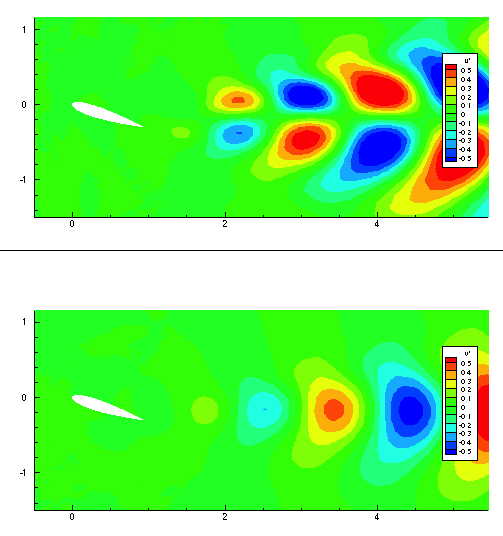}}
\caption{Least stable mode for NACA0015 stability studies. Top: $u'$ contours. Bottom: $v'$ contours.}
\label{fig:naca_mode1}
\end{figure}
All eigenmodes are found to be decaying for both cases, i.e., they are stable. However, the most weakly damped modes occur at different frequencies
For $M=0.1$, the least stable mode is found to be at a frequency of $\omega=0.384$,  which is very close to the $\omega \simeq 0.38$ value reported by \citet{he2017linear} using Arnoldi based time-stepping approach. 
However, as the Mach number is increased to $M=0.5$, the frequency of the 
most weakly damped 
mode decreases to $\omega=0.2893$. 
This decrease in frequency with increase in Mach number is also observed  in cylinder wake study of \citet{canuto2015two}.
However,  the current reduction of around 24\% is more drastic than the 9\% obtained between $M=0$ and $M=0.5$ near the critical $Re$ for the cylinder in their study. 
\citet{canuto2015two} also observe that compressibility reduces the vortex-shedding frequency while increasing the wavelength, and but this effect is less pronounced as $Re$ increases towards and beyond the critical value.
Thus, the current higher reduction in frequency may be attributed to the fact that Reynolds number is far below critical.    

The spatial structures of the least stable modes are shown in Fig.~\ref{fig:naca_mode1}.
In both Mach number cases, these modes are similar to Kelvin-Helmholtz modes observed in the wake of a bluff body such as a circular cylinder. 
 The structures are more prominent at $M=0.1$ than at $M=0.5$.
 The former is thus more susceptible to instability than the latter, for which compressibility acts as a stabilizing agent by countering the vorticity generation through mechanisms such as vorticity dilation as well as baroclinic torque \cite{Ohmichi2017compressibility}.  
 A more detailed assessment reveals that the wavelength of the modes increases with Mach number in consistent with the findings of \citet{canuto2015two}. 
 This effect is due to compressibility, which is responsible for the elongation of the recirculation region as shown in Fig. \ref{fig:naca_mean}.  
The test case indicates the manner in which the current approach easily assimilates the treatment of such flows in a compressible framework.

\section{Concluding remarks}\label{sec:conclusion}
A generalizable and parameter-free approach to obtain stability modes is presented.
The method falls into the class of time-stepping or time-stepper methods.
The required Jacobian-vector product subspace is efficiently generated  by  simple modifications to existing non-linear solvers, thus leveraging mature existing three-dimensional higher order spatio-temporal discretization techniques to ensure implicit linearization without the need for Fre\`{c}het derivatives.
The initial vector contains the spatio-temporal scales of interest through random impulse perturbations.
The accurate subspace satisfies the requirements of dynamic mode decomposition (DMD) for stability mode extraction .
When DMD is applied in a post processing manner to the subspace, the eigenmodes are the primary and other dominant stability modes, which are obtained in a straightforward and iteration-free manner. This is particularly advantageous in flows with two or more competing mechanisms that determine stability.
Mode convergence tests are easily established by varying the size and identity of the subspace.  
The results indicate that snapshots based on only few relevant variables, such as velocity or vorticity, are sufficient to extract stability mode features of interest.
This yields a substantial reduction in memory requirements of upto $80\%$, which enables more routine treatment of multi-dimensional systems. 
Steps such as orthonormalization, preconditioning, operator inversion or specification of initial guesses are avoided. 
Application to a variety of incompressible and compressible cases, including benchmark flows as well as those requiring curvilinear coordinates, demonstrate the accuracy, efficiency and robustness of the method.
A detailed comparison of current approach with Arnoldi method for a compressible lid-driven cavity show that while the accuracy of the results are comparable, former has considerable advantages in terms of requirement of the computational resources and other practical considerations.       

\section*{References}\label{sec:ref}
\bibliographystyle{model1-num-names}
\bibliography{refer}
\end{document}